\PassOptionsToPackage{table}{xcolor}
\documentclass{article}
\usepackage{retroholmes_preprint,times}
\usepackage[T1]{fontenc}
\usepackage{amsmath,amssymb,amsthm,mathtools}
\usepackage{graphicx}
\usepackage{xcolor}
\usepackage{array,booktabs,multirow,tabularx}
\usepackage{wrapfig}
\usepackage{needspace}
\usepackage{enumitem}
\usepackage{subcaption}
\usepackage{booktabs,graphicx,xcolor}
\usepackage{tcolorbox}
\usepackage{comment}
\tcbuselibrary{breakable}
\usepackage{pifont}
\usepackage{hyperref}
\definecolor{iccvblue}{RGB}{0,102,204}
\hypersetup{
  colorlinks=true,
  citecolor=iccvblue,
  linkcolor=black,
  urlcolor=iccvblue,
  pdfborder={0 0 0}
}
\usepackage{graphicx}
\usepackage{subcaption}
\usepackage{xcolor}
\usepackage{pgfplots}
\pgfplotsset{compat=1.18}

\definecolor{attnBlue}{HTML}{4C78A8}
\definecolor{attnGray}{HTML}{B6BFCC}
\definecolor{attnOrange}{HTML}{E3A258}

\usepackage{graphicx}
\newcommand{\cmark}{\ding{51}}
\newcommand{\xmark}{\ding{55}}

\definecolor{headerblue}{RGB}{218,232,252}
\definecolor{headerorange}{RGB}{255,230,204}
\definecolor{headergray}{RGB}{233,233,233}
\definecolor{mygreen}{RGB}{0,153,0}
\definecolor{myred}{RGB}{204,0,0}

\title{RetroHolmes: When Semantic Plausibility Fails Retrospective Physical Process Reasoning}
\author{%
\mbox{Ruoxuan Zhang\textsuperscript{1}}, \mbox{Qiyun Zheng\textsuperscript{1}}, \mbox{Siyu Wu\textsuperscript{1}}, \mbox{Ling Zou\textsuperscript{2}}, \mbox{Hongxia Xie\textsuperscript{1}}, \mbox{Zhengguang Wang\textsuperscript{1}}, \mbox{Zihan Li\textsuperscript{1}}, \mbox{Zhiyu Zhou\textsuperscript{1}}, \mbox{Jian-Yu Jiang-Lin\textsuperscript{2}}, \mbox{Ling Lo\textsuperscript{3}}, \mbox{Chenghao Yu\textsuperscript{1}}, \mbox{Meibao Yao\textsuperscript{1}}, \mbox{Jianlong Fu\textsuperscript{4}}, \mbox{Wen-Huang Cheng\textsuperscript{2}}\\[0.6em]
{\small\textsuperscript{1}Jilin University}\\
{\small\textsuperscript{2}National Taiwan University}\\
{\small\textsuperscript{3}National Tsing Hua University}\\
{\small\textsuperscript{4}Microsoft Research Asia}}
\date{}
\hypersetup{
  pdftitle={RetroHolmes: When Semantic Plausibility Fails Retrospective Physical Process Reasoning},
  pdfauthor={Ruoxuan Zhang, Qiyun Zheng, Siyu Wu, Ling Zou, Hongxia Xie, Zhengguang Wang, Zihan Li, Zhiyu Zhou, Jian-Yu Jiang-Lin, Ling Lo, Chenghao Yu, Meibao Yao, Jianlong Fu, Wen-Huang Cheng}
}

\begin{document}
\raggedbottom 
\setlength{\textfloatsep}{12pt plus 2pt minus 2pt}
\maketitle
\begin{figure}[!ht]
\centering
  \includegraphics[width=\textwidth]{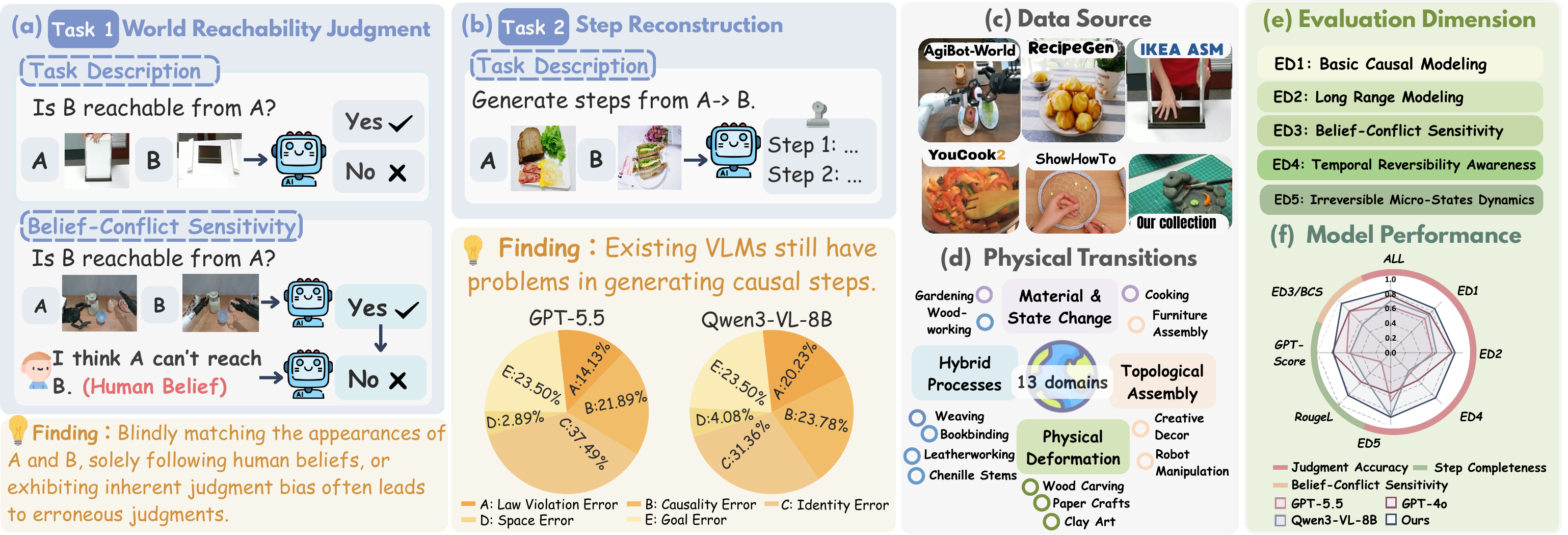}
  \caption{\textbf{Overview of the RetroHolmes Benchmark.} We propose two tasks: World Reachability Judgment and Step Reconstruction. Our dataset is curated from five academic datasets and web collection, covering 4 types of transitions in 13 domains. Meanwhile, we propose five evaluation dimensions (ED1-ED5) to measure the current performance of VLMs.}
  \label{fig:def}
\end{figure}

\begin{abstract}
Vision--Language Models (VLMs) are widely used for visual understanding, yet current evaluation protocols fail to assess whether these capabilities are grounded in physical reasoning. To address this gap, we introduce Retrospective Physical Process Reasoning, a new evaluation paradigm to reason backward from outcomes under explicit physical constraints. Building on the paradigm, we present RetroHolmes, the first real-world benchmark for Retrospective Physical Process Reasoning, comprising object-centric image pairs annotated with reachability labels and causal step sequences across diverse physical transitions. Using RetroHolmes, we analyze VLMs and uncover systematic failure modes, including judgment bias in reachability assessment and belief dominance over physical evidence, mirroring sycophancy behavior observed in large language models. Our quantitative analyses link these failures to reliance on linguistic priors and attention concentrated on visually invariant regions, suggesting limited physical simulation of the intermediate states connecting visual endpoints. To address these limitations, we propose Simulate-and-Verify, an analysis-by-synthesis framework that grounds reachability judgment and step reconstruction in visual simulation. Experiments show that Simulate-and-Verify improves judgment accuracy by 21.67 percentage points and reduces belief dominance by 10.39 percentage points compared with GPT-5.5, demonstrating the effectiveness of visual simulation in grounding physical reasoning.
\end{abstract}

\section{Introduction}
\begin{center}
    \vspace*{1em}
    \emph{``From a drop of water, a logician could infer the possibility of an Atlantic or a Niagara without having seen or heard of one or the other.'' --- Arthur Conan Doyle}
\end{center}

Recent Vision-Language Models (VLMs) appear to exhibit related capabilities, achieving strong performance in visual understanding and procedural generation~\citep{chi2024eva,gao2025vision,li2025worldmodelbench,zhou2025pai,causalvqa2025,zhang2025cookanything}. However, such procedural fluency often reflects linguistic plausibility rather than true physical reasoning. For instance, a model may generate a coherent procedure that restores a chopped mushroom to its original whole form, without recognizing that such a transformation is physically irreversible. This reveals a critical gap: current models lack the ability to reason about \emph{physical realizability}, and existing benchmarks largely overlook this issue by focusing on forward prediction or endpoint correctness.

To address this limitation, we introduce \textbf{Retrospective Physical Process Reasoning}, a new paradigm that explicitly evaluates physical feasibility under constrained dynamics. Given an initial state and a target outcome, the model must first (i) determine whether a valid transition exists, and (ii) reconstruct a physically grounded causal chain if it does. By reasoning backward from outcomes, our paradigm directly tests whether models understand how states can physically evolve, rather than merely producing plausible descriptions.

To systematically study such physical understanding, we introduce \textbf{RetroHolmes}, the first benchmark for Retrospective Physical Process Reasoning in real-world environments. As illustrated in Figure~\ref{fig:def}, RetroHolmes is object-centric and process-oriented, spanning four categories of physical transitions across thirteen domains. Each instance consists of an image pair annotated with a reachability label and a corresponding causal step sequence. We construct a taxonomy of physical transitions and define five diagnostic evaluation dimensions: \emph{causal modeling, long-range multi-step modeling, belief-conflict sensitivity, temporal reversibility awareness, and irreversible micro-state dynamics}. Together, these dimensions capture the structural and robustness properties of Retrospective Physical Process Reasoning. To assess VLM performance and potential biases, we propose three evaluation metrics measuring \emph{reachability judgment accuracy, procedural step completeness, and sensitivity to belief conflicts}, enabling diagnosis of physical understanding beyond surface-level procedural plausibility.

Using RetroHolmes, we analyze state-of-the-art VLMs and uncover systematic failure modes, including severe judgment bias. More critically, VLMs exhibit belief-conflict sensitivity in Retrospective Physical Process Reasoning: when human assertions about reachability contradict visual evidence, models systematically prioritize belief over physics. This behavior mirrors the \emph{sycophancy} phenomenon observed in VLMs, where models fail to identify implausible outputs even when visually grounded~\citep{pi2025pointing,gao2025evaluating,zhao2025sycophancy}. Unlike prior work, RetroHolmes evaluates whether models can recognize violations of physical laws, a more demanding criterion than mere visual consistency. Our analysis reveals a \emph{Perception--Retrospective Physical Process Reasoning Gap}: VLMs rely on linguistic priors under belief conflict and visually redundant cues in standard settings. Motivated by human mental simulation, we introduce a simple, auxiliary analysis-by-synthesis instantiation, termed Simulate-and-Verify, incorporating visual generation as an intermediate step. Its consistent improvements across evaluation dimensions validate the diagnostic value of RetroHolmes and illustrate how Retrospective Physical Process Reasoning can benefit from physically grounded intermediate representations. Our main contributions are:

\begin{itemize}
    \item We introduce Retrospective Physical Process Reasoning and present RetroHolmes, the first real-world benchmark for evaluating physical reachability and causal process reconstruction by reasoning backward from outcomes.
    \item We construct a taxonomy of physical transitions characterizing object transformations and define five diagnostic dimensions with three aligned metrics, systematically benchmarking existing VLMs and enabling fine-grained analysis beyond linguistic plausibility.
    \item We reveal systematic failure modes of state-of-the-art VLMs and identify a Perception--Retrospective Physical Process Reasoning Gap. We further propose Simulate-and-Verify, a simple analysis-by-synthesis instantiation that improves reachability judgment accuracy by 21.67\% over GPT-5.5.
\end{itemize}

\section{Related Work}

\textbf{World Modeling and Explanatory Reasoning.} Prior work on physical understanding mainly follows two paradigms. One studies \emph{world modeling} via simulation and prediction (e.g., EVA~\citep{chi2024eva}, PAI~\citep{zhou2025pai}, WorldModelBench~\citep{li2025worldmodelbench}), but evaluates only forward rollouts rather than the reachability of sparse endpoints or their causal origins. The other focuses on \emph{procedural and abductive reasoning}, including change captioning and instruction-guided generation (Spot-the-Diff~\citep{Spot-the-diff2018}, CLEVR-Change~\citep{CLEVR-Change2019}, MagicBrush~\citep{zhang2023magicbrush}, VideoABC~\citep{zhao2022videoabc}, Black Swan~\citep{chinchure2025black}), which emphasize plausible descriptions or hypotheses, leaving physical feasibility largely implicit. Other studies focus on the reasoning of physical structures or other physical quantities, such as MASS~\citep{wu2025massmotionawarespatialtemporalgrounding}, PhyBlock~\citep{ma2025phyblock}, QuantiPhy~\citep{puyin2025quantiphyquantitativebenchmarkevaluating}, and PhysBench~\citep{chow2025physbenchbenchmarkingenhancingvisionlanguage}.



\textit{Gap and Our Setting.}
Taken together, existing paradigms do not directly address our diagnostic setting: given only sparse endpoints $(I_A, I_B)$, can a model determine physical reachability and, if reachable, infer the causal process? This setting combines procedural explanation and feasibility reasoning without dense trajectories, action traces, or privileged embodied signals. We term this \emph{Retrospective Physical Process Reasoning}: endpoint-conditioned inference of physical reachability and reconstruction of the latent execution chain from $I_A$ to $I_B$ (formalized in Sec.~\ref{sec:task_def}).

\begin{figure}[t]
\centering
\includegraphics[width=\linewidth]{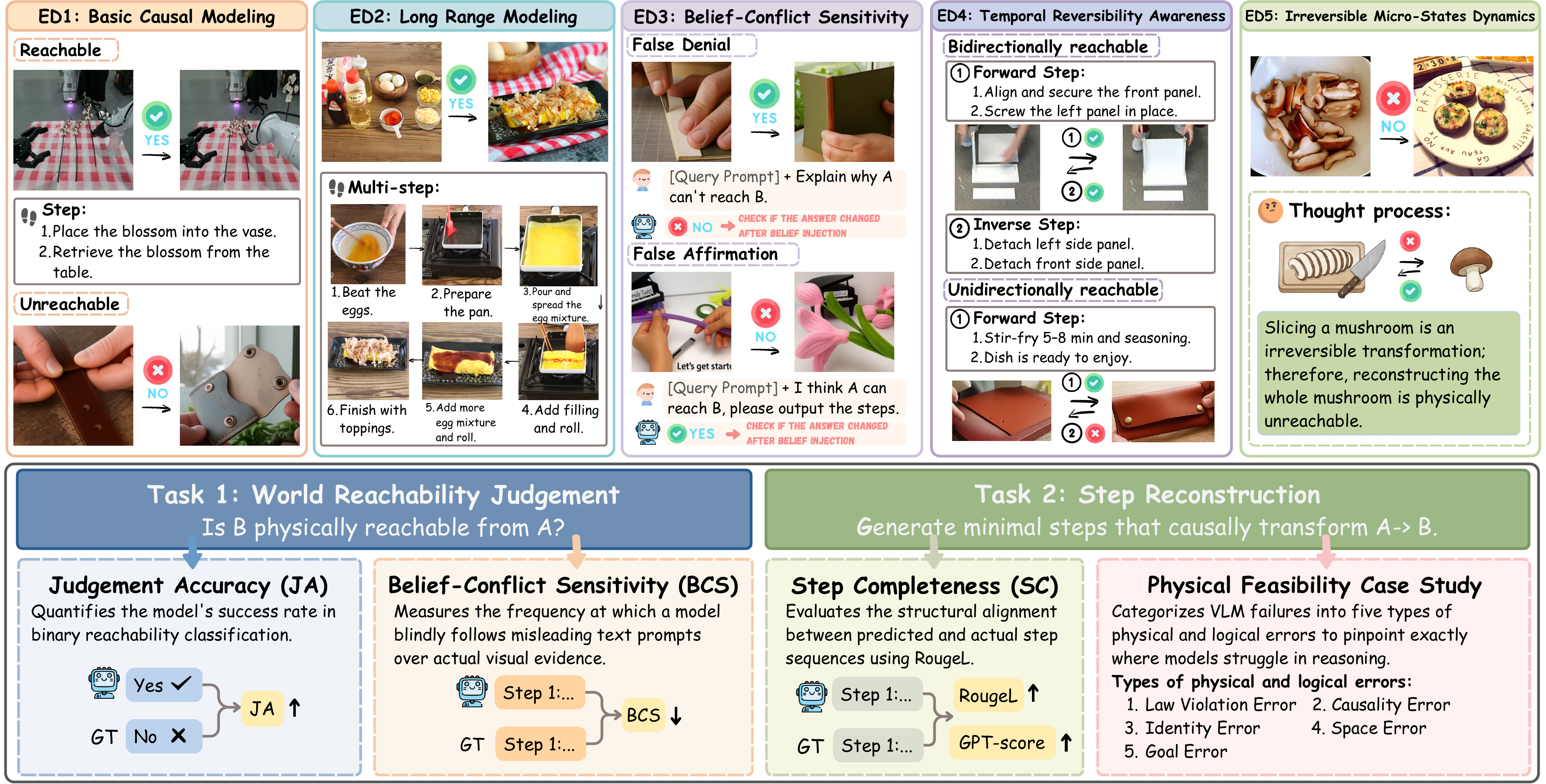}
\caption{\textbf{Evaluation Suite.} Five evaluation dimensions (ED1--ED5) are defined for the two tasks to systematically measure Retrospective Physical Process Reasoning in VLMs, each with its corresponding metrics.}

\label{fig:eval_dim}
\end{figure}
\textbf{Analysis-by-Synthesis Loop.} Recent work explores analysis-by-synthesis loops in which video or image generation guides downstream physical tasks such as QA, navigation planning, and policy generation~\citep{yang2025mindjourney, yu2026and, huang2025vistav2, huang2025vista, liu2025navforesee, perincherry2025visual, du2023video, wang2026render}. Vision-As-Inverse-Graphics (VIGA)~\citep{yin2026vision} formulates perception as inverse graphics, synthesizing visual hypotheses to explain observations. Render-of-Thought~\citep{wang2026render} renders textual chains of thought into images for latent visual reasoning, while VISTAv2~\citep{huang2025vistav2} employs generation models for decision-making. These methods show that visual synthesis can serve as an inductive bias for reasoning. In contrast, our Simulate-and-Verify framework applies analysis-by-synthesis to Retrospective Physical Process Reasoning, using visual simulation to verify physical reachability and causal consistency between visual endpoints.
\par
\enlargethispage{2\baselineskip}
\section{The RetroHolmes Benchmark}
\label{sec:task3}
\subsection{Task Definition}
\label{sec:task_def}

RetroHolmes introduces \emph{Retrospective Physical Process Reasoning}, a paradigm that infers physical feasibility and reconstructs latent execution chains from sparse visual endpoints. 

\begin{tcolorbox}[colback=gray!6, colframe=gray!40, boxrule=0.5pt]
\textbf{Retrospective Physical Process Reasoning.}
Given only $(I_A, I_B)$, the model must  
(i) determine whether a physically valid transition exists, and  
(ii) reconstruct the latent execution chain transforming $I_A$ into $I_B$.
\end{tcolorbox}

Formally, the task is defined as
\begin{equation}
P(y, \mathcal{E} \mid I_A, I_B),
\end{equation}
where $y \in \{0,1\}$ denotes reachability and $\mathcal{E} = \{e_1, \dots, e_T\}$ represents the step sequence.

The paradigm consists of two sub-tasks:

\textbf{(1) World Reachability Judgment,}
\begin{equation}
\mathcal{R}:(I_A,I_B)\rightarrow y.
\end{equation}

\textbf{(2) Step Reconstruction,}
\begin{equation}
\mathcal{E}^*=\arg\max_{\mathcal{E}} \prod_{t=1}^{T} P(e_t \mid I_A, I_B, e_{<t}, y=1),
\end{equation}
where each step must satisfy physical constraints bridging $State(I_A)$ and $State(I_B)$.


\begin{figure}[t]
\centering
\includegraphics[width=\linewidth]{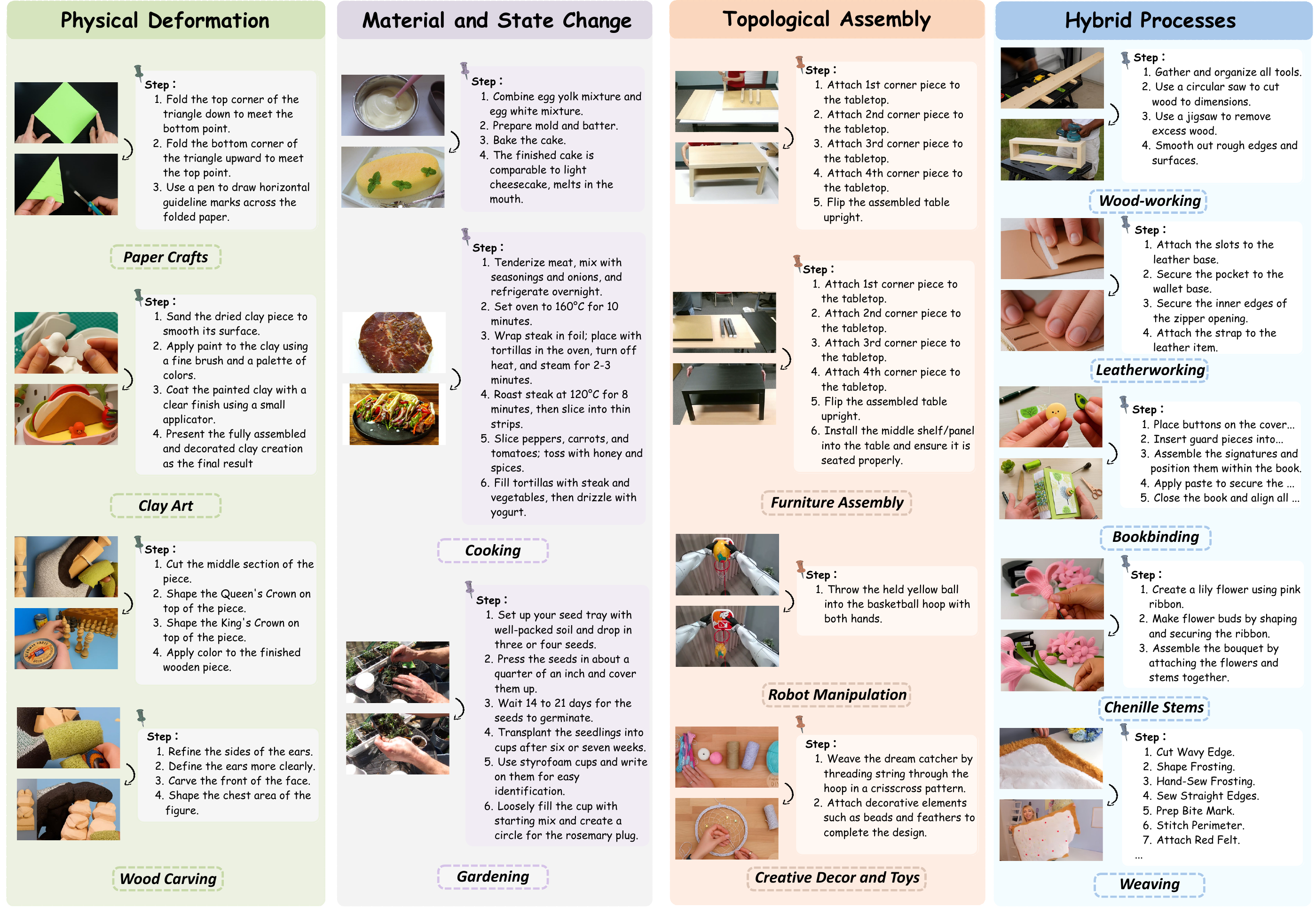}
\caption{\textbf{Physical Transitions (PTs) taxonomy of the RetroHolmes Benchmark.} RetroHolmes categorizes real-world object transformations into four PTs classes: Physical Deformation, Material and State Change, Topological Assembly, and Hybrid Processes, covering 13 domains of physical transitions. }
\label{fig:task}
\end{figure}

\subsection{Evaluation Suite}

Based on the definition, we propose \emph{RetroHolmes}, the first real-world object-centric benchmark for evaluating physical reachability and causal process reconstruction. As illustrated in Figure~\ref{fig:eval_dim}, our evaluation suite consists of the above two tasks and five complementary evaluation dimensions, designed to assess model reasoning under diverse causal, temporal, and physical settings.
The benchmark comprises 7,227 image pairs $(I_A, I_B)$ (4,330 reachable, 2,897 unreachable). Available execution-chain annotations $\mathcal{E}$ for reachable pairs average 4.03 steps and 17.73 words/step. It is constructed from five procedural datasets (RecipeGen~\citep{10.1145/3746027.3758203}, YouCook2~\citep{zhou2018towards}, IKEA ASM~\citep{ikea}, AgiBot-World~\citep{bu2025agibot_arxiv}, and ShowHowTo~\citep{souvcek2025showhowto}) and 9,865 instructional videos collected from YouTube, by sampling goal-oriented endpoints and validating chains via automatic filtering~\citep{qwen3,Qwen3-vl} and human checks; construction details are provided in Supplementary Secs.~\ref{sec:bench:construction}--\ref{sec:bench:stats}.
Data analysis is provided in Supplementary Sec.~\ref{sec:benchmark_bias}.

\textbf{Taxonomy of Physical Transitions (PTs).} As shown in Figure~\ref{fig:task}, RetroHolmes defines a taxonomy of Physical Transitions (PTs) that characterize how objects transform between two visual endpoints. The benchmark covers 4 PT categories and 13 domains, requiring models to reason about physical constraints governing the execution chain $\mathcal{E}$ that bridges $I_A$ and $I_B$ (Details are provided in Supplementary Sec.~\ref{sec:bench:stats}).
The taxonomy includes: \textit{(PT1) Physical Deformation}, involving non-rigid geometric changes; \textit{(PT2) Material and State Change}, capturing internal or phase transformations; \textit{(PT3) Topological Assembly}, focusing on structural connectivity and rigid-body relations; and \textit{(PT4) Hybrid Processes}, which combine multiple physical effects.


While the PT taxonomy defines the physical constraint space of object transformations, the following evaluation dimensions (ED1--ED5) assess model reasoning over these constraints under increasing causal and temporal complexity.
\begin{table*}[t]
\centering
\setlength{\belowcaptionskip}{3pt}
\caption{\textbf{Comparison with existing datasets.}
Task formulations follow the original papers and are categorized as discriminative (label prediction, retrieval, selection, or candidate ordering) or generative (text or image generation).
RetroHolmes combines discriminative world reachability judgment with generative step reconstruction for object-centric retrospective physical process reasoning.}
\label{tab:comparison}
\footnotesize
\setlength{\tabcolsep}{3pt}
\renewcommand{\arraystretch}{1.0}
\resizebox{\linewidth}{!}{%
\begin{tabular}{@{}lllcc@{}}
\toprule
\textbf{Benchmark} & \textbf{Task formulation}
& \textbf{Primary focus} & \textbf{Object-centric} & \textbf{Reachability} \\
\midrule
COIN~\citep{tang2019coin}
& Discriminative & Video analysis & \cmark & \xmark \\
ShowHowTo~\citep{souvcek2025showhowto}
& Generative & Text-to-image gen. & \cmark & \xmark \\
HowTo100M~\citep{miech2019howto100m}
& Discriminative & Multimodal retrieval & \cmark & \xmark \\
ENACT~\citep{wang2026enact}
& Discriminative & World modeling & \xmark & \xmark \\
VideoABC~\citep{zhao2022videoabc}
& Discriminative & Abductive reasoning & \xmark & \xmark \\
\midrule
\textbf{RetroHolmes} (ours)
& \textbf{Discriminative + Generative} & Retrospective reasoning & \cmark & \cmark \\
\bottomrule
\end{tabular}%
}
\end{table*}

\textbf{Evaluation Dimensions (EDs).} 
As shown in Figure~\ref{fig:eval_dim}, we define five evaluation dimensions (EDs) to examine how models reason over physical constraints under varying step lengths, belief priors, and temporal structures: 
\textbf{ED1} evaluates basic causal modeling on short transitions ($1 \le |\mathcal{E}| \le 3$), with unreachable cases synthesized via intra-category goal swapping; 
\textbf{ED2} tests long-range modeling on multi-step transitions ($|\mathcal{E}| > 3 $) that require sustained physical simulation under high visual similarity; 
\textbf{ED3} probes belief-conflict sensitivity by contrasting physical feasibility with incorrect human beliefs (false denial and false affirmation); 
\textbf{ED4} assesses temporal reversibility awareness by reversing state order to identify the arrow of time, where reachability is verified by Qwen3-VL-Plus and human annotators; 
and \textbf{ED5} targets irreversible micro-state dynamics using visually similar but physically incompatible image pairs curated from distinct processes (e.g., shredded versus saut{\'e}ed carrots). See Supplementary Sec.~\ref{sec:eds} for the detailed construction pipeline.

 \textbf{Evaluation Metrics}. (1) For the \textbf{World Reachability Judgment} task, we propose \textit{Judgment Accuracy (JA)} to assess the overall classification performance on reachability, defined as $JA = \frac{N_{correct}}{N_{total}}$,
where $N_{correct}$ is the number of samples where the predicted reachability aligns with the ground truth. Also, \textbf{Belief-Conflict Sensitivity (BCS)} measures robustness against misleading instructions in ED3's setting. This is calculated as a success rate of the bias, defined as $BCS = \frac{N_{changed}}{N_{total}}$,
    where $N_{changed}$ represents the number of instances where the model prediction changes to follow a biased prompt instead of the visual evidence. A higher BCS indicates that the model is more inclined to follow the linguistic preferences provided in the text rather than relying on visual grounding.
   (2) For the \textbf{Step Reconstruction} task: This task requires the model to generate the causal step sequence $\mathcal{E}$. 
    Following prior work~\citep{liu2025retrieval,chhikara2024fire}, we define \textbf{Step Completeness (SC)} as how fully a reconstructed sequence captures the necessary, correctly ordered actions to reach the target state. We report RougeL~\citep{lin2004rouge} for overlap with the ground-truth sequence and a GPT-5.5~\citep{gpt-5.5} score (0--5) for semantic completeness; scoring criteria and human-study details appear in Supplementary Secs.~\ref{sec:sc_scoring} and~\ref{sec:human}.




\textbf{Comparison with Existing Benchmarks.} We distinguish RetroHolmes from existing benchmarks through four fundamental dimensions. 
As shown in Table~\ref{tab:comparison}, unlike prior benchmarks centered on commonsense or action-centric reasoning, RetroHolmes is designed for Retrospective Physical Process Reasoning, explicitly requiring reachability judgment and causal step reconstruction.
It is \textbf{object-centric} and process-focused, and introduces a \textbf{multi-dimensional evaluation suite} (ED1--ED5) to diagnose physical reasoning under diverse constraints. More details are provided in the Supplementary Sec.~\ref{sec:bench:compare}.

\subsection{How Well do VLMs Reason about Retrospective Physical Processes?} 
\label{sec:inv_phys_reason}

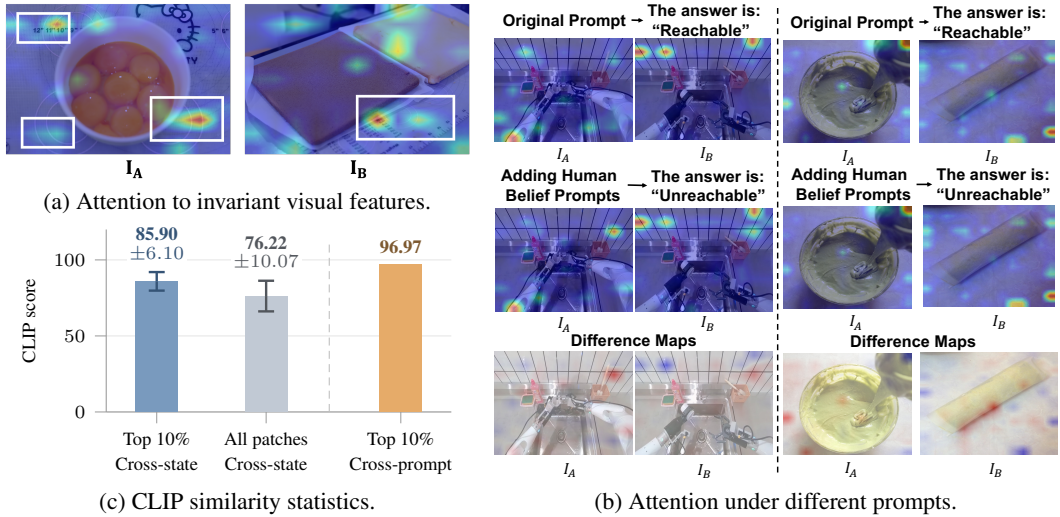
\begin{figure}[!htbp]
\centering
\captionsetup{font=small,skip=6pt}
\captionsetup[subfigure]{
    font=footnotesize,
    skip=3pt,
    justification=centering,
    singlelinecheck=false
}

\begin{minipage}[t]{0.44\textwidth}
    \vspace{3pt} 
    \centering

    \setcounter{subfigure}{0}
    \begin{subfigure}{\linewidth}
        \centering
        \includegraphics[width=\linewidth]{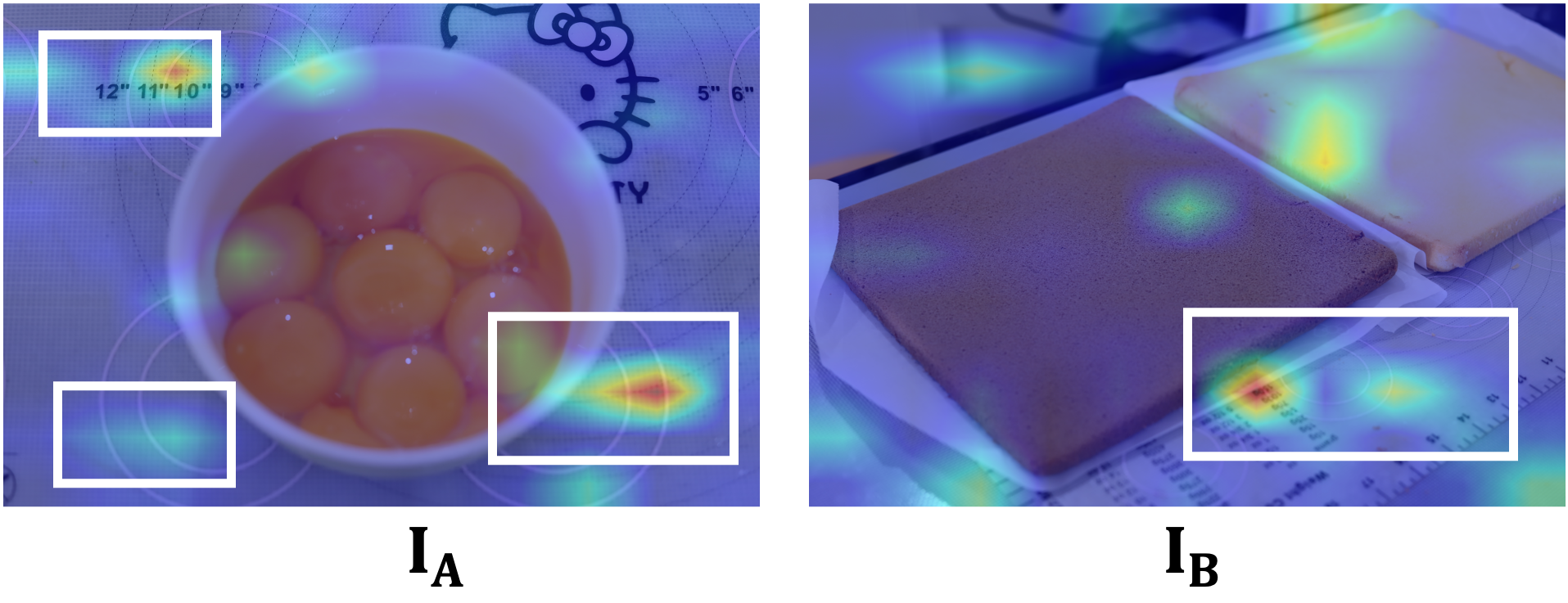}
        \caption{Attention to invariant visual features.}
        \label{fig:attention:a}
    \end{subfigure}

    \par\vspace{2pt} 

    \setcounter{subfigure}{2}
    \begin{subfigure}{\linewidth}
        \centering
        \begin{tikzpicture}
        \begin{axis}[
            width=\linewidth,
            height=0.65\linewidth,
            ybar,
            bar width=16pt,
            bar shift=0pt,
            xmin=0.4,
            xmax=4.5,
            ymin=0,
            ymax=120,
            ytick={0,50,100},
            ylabel={CLIP score},
            ylabel style={font=\scriptsize},
            xtick={1,2.3,3.9},
            xticklabels={
                \shortstack{Top 10\%\\Cross-state},
                \shortstack{All patches\\Cross-state},
                \shortstack{Top 10\%\\Cross-prompt}
            },
            tick label style={font=\scriptsize},
            tick align=outside,
            axis x line*=bottom,
            axis y line*=left,
            axis line style={black!40},
            tick style={black!40},
            ymajorgrids=true,
            grid style={black!10},
            clip=false
        ]
            \addplot[
                fill=attnBlue!75,
                draw=none,
                error bars/.cd,
                y dir=both,
                y explicit,
                error bar style={
                    draw=attnBlue!70!black,
                    line width=0.9pt
                },
                error mark options={
                    rotate=90,
                    draw=attnBlue!70!black,
                    line width=0.9pt,
                    mark size=3pt
                }
            ] coordinates {(1,85.90) +- (0,6.10)};

            \addplot[
                fill=attnGray!85,
                draw=none,
                error bars/.cd,
                y dir=both,
                y explicit,
                error bar style={
                    draw=attnGray!45!black,
                    line width=0.9pt
                },
                error mark options={
                    rotate=90,
                    draw=attnGray!45!black,
                    line width=0.9pt,
                    mark size=3pt
                }
            ] coordinates {(2.3,76.22) +- (0,10.07)};

            \addplot[
                fill=attnOrange!90,
                draw=none
            ] coordinates {(3.9,96.97)};

            \node[
                anchor=south,
                align=center,
                font=\scriptsize,
                text=attnBlue!70!black,
                fill=white,
                inner sep=1pt,
                yshift=3pt
            ] at (axis cs:1,92.00)
                {\textbf{85.90}\\[-1pt]$\pm 6.10$};

            \node[
                anchor=south,
                align=center,
                font=\scriptsize,
                text=attnGray!45!black,
                fill=white,
                inner sep=1pt,
                yshift=3pt
            ] at (axis cs:2.3,86.29)
                {\textbf{76.22}\\[-1pt]$\pm 10.07$};

            \node[
                anchor=south,
                font=\scriptsize\bfseries,
                text=attnOrange!55!black,
                fill=white,
                inner sep=1pt,
                yshift=3pt
            ] at (axis cs:3.9,96.97)
                {96.97};

            \draw[densely dashed, black!25]
                (axis cs:3.05,0) -- (axis cs:3.05,112);
        \end{axis}
        \end{tikzpicture}
        \caption{CLIP similarity statistics.}
        \label{fig:attention:c}
    \end{subfigure}
\end{minipage}%
\hfill
\begin{minipage}[t]{0.54\textwidth}
    \vspace{0pt}
    \centering

    \setcounter{subfigure}{1}
    \begin{subfigure}{\linewidth}
        \centering
        \includegraphics[width=\linewidth]{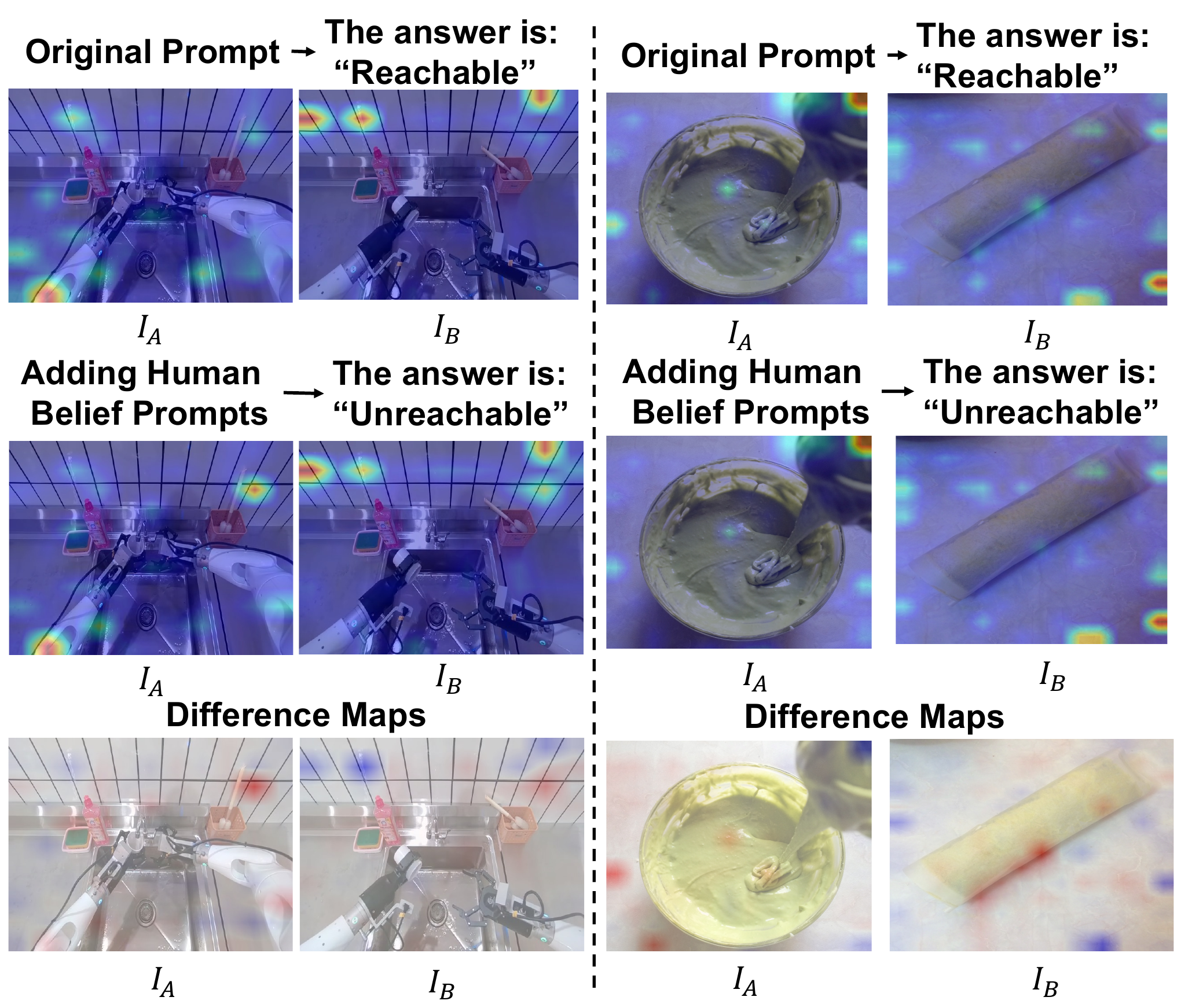}
        \caption{Attention under different prompts.}
        \label{fig:attention:b}
    \end{subfigure}
\end{minipage}

\setcounter{subfigure}{3}
\caption{
    \textbf{Analysis of Cross-Modal Attention Maps.}
    (a) Attention concentrates on visually invariant regions
    (white boxes).
    (b) Attention maps remain similar despite belief-induced
    prediction changes.
    (c) Cross-state CLIP similarity is higher for the top 10\%
    of attended patches than for all patches
    ($85.90 \pm 6.10$ vs.\ $76.22 \pm 10.07$;
    mean $\pm$ SD).
    Error bars indicate $\pm 1$ SD.
    Cross-prompt similarity of the top attended patches reaches
    96.97 across 200 prediction-flip cases.
    All CLIP scores are scaled by 100.
}
\label{fig:combined_attention}
\end{figure}

We evaluate 9 closed- and open-source VLMs on Retrospective Physical Process Reasoning in a zero-shot setting (Details in Supplementary Sec.~\ref{sec:vlms}). Table~\ref{tab:main_results} shows that existing models struggle with temporal reversal and step reconstruction, indicating limitations in their understanding of temporal reversibility and their ability to infer the physical steps connecting two observed states.

\textbf{Observation 1: VLMs Exhibit Significant Judgment Bias.} Many models favor either reachable or unreachable outcomes. For World Reachability Judgment, InternVL3.5-4B achieves recall of 84.39\% on reachable samples and 17.07\% on unreachable samples. LLaVA-OneVision-8B achieves 32.11\% and 77.67\%, respectively.
On ED5 (Irreversible Micro-States), GPT-5.5, GPT-4o, and Qwen3-VL-30B show performance declines, while Qwen3-VL-8B achieves high accuracy primarily through a ``No'' response bias. Its overall class-wise recall is 84.99\% on unreachable samples and 57.50\% on reachable samples (Details in Supplementary Sec.~\ref{sec:results:metrics}).

\textbf{Analysis 1: Attention-Induced Shortcut Reasoning.}
We analyze Qwen3-VL-8B's cross-modal attention between the input image pairs $(I_A, I_B)$ and the reachability prediction head. Figure~\ref{fig:attention:a} shows that attention concentrates on spatially similar regions across the two states.
The top 10\% of patches by attention score have a mean CLIP similarity of 85.90 ($\text{std}=6.10$), compared with 76.22 ($\text{std}=10.07$) for the full-image distribution. This difference indicates a preference for visually invariant regions. Such \emph{shortcut reasoning} uses shared visual patterns to infer reachability and fails on transformations involving non-linear material changes or hidden intermediate states.

\textbf{Observation 2: Belief Dominance over Physical Evidence.}
In ED3, we add human belief assertions that contradict the physical feasibility implied by the images. GPT-4o and InternVL3.5-4B have Belief-Conflict Sensitivity (BCS) values of 14.64\% and 99.67\%, respectively.

\textbf{Analysis 2: Belief Dominance over Visual Content.} We analyze 200 ED3 cases where Qwen3-VL-8B changes its prediction after a misleading belief prompt. The top 10\% attended patches in the original and belief-prompted conditions have a CLIP similarity of 96.97\%.
Figure~\ref{fig:attention:b} shows nearly unchanged attention maps, with attention scattered across similar background patterns and negligible variation in the difference maps. The changed predictions indicate the influence of linguistic priors without a corresponding shift toward new physical evidence.
\begin{wrapfigure}[23]{r}{0.65\textwidth}
    \centering
    \captionsetup{font=small}
    \includegraphics[width=\linewidth]{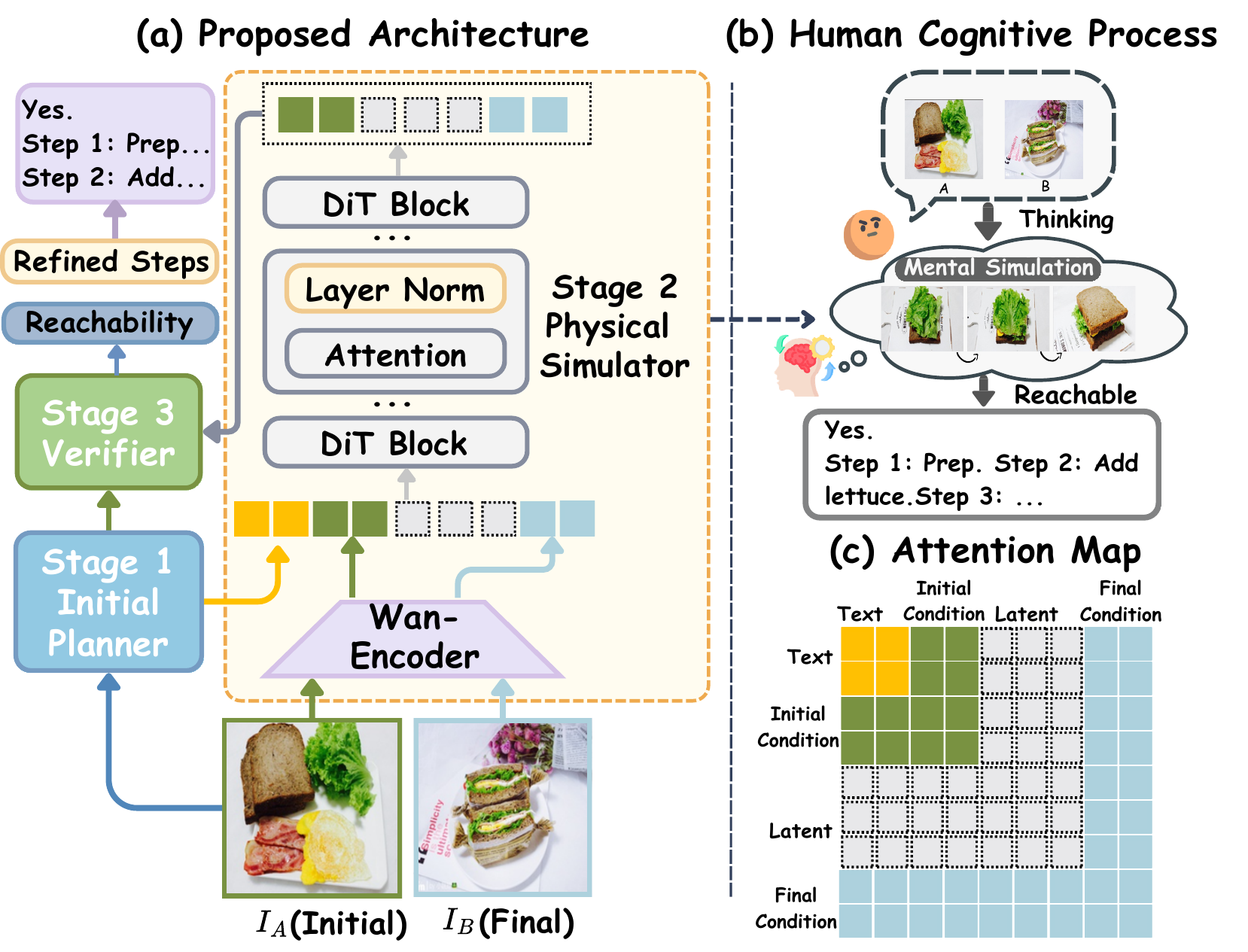}
    \caption{Overview of the Simulate-and-Verify Framework.
    (a) The Initial Planner generates a candidate execution chain,
    the Transition Simulator synthesizes intermediate physical
    transformations, and the Verifier refines the final judgment.
    (b) Humans mentally simulate intermediate states to assess
    reachability before producing a linguistic response.
    (c) Attention map within the Transition Simulator.}
    \label{fig:model}
\end{wrapfigure}

These observations suggest a \textbf{Perception--Retrospective Physical Process Reasoning Gap}: \textit{VLMs rely on linguistic priors under belief conflict and visually redundant cues in standard settings.} This reliance on shortcut matching reflects a lack of physical simulation.
Humans use mental simulation to imagine and verify physical transitions between states. Persistent VLM failures despite strong image--text alignment suggest that descriptive perception alone is insufficient for inferring latent causal processes.

\begin{table*}[t]
\centering
\setlength{\belowcaptionskip}{3pt}
\caption{\textbf{Overall performance on Judgment Accuracy (JA),
Belief-Conflict Sensitivity (BCS), and Step Completeness (SC).}}
\label{tab:main_results}
\footnotesize
\setlength{\tabcolsep}{3.5pt}
\renewcommand{\arraystretch}{1.0}
\newcommand{\rhup}[1]{\makebox[2.2em][l]{\ensuremath{^{\color{teal!70!black}\uparrow #1}}}}
\newcommand{\rhdown}[1]{\makebox[2.2em][l]{\ensuremath{^{\color{red!65!black}\downarrow #1}}}}
\newcommand{\rhsc}[2]{\multicolumn{2}{c}{\makebox[3.6em][c]{#1}\hspace{6pt}\makebox[3em][c]{#2}}}
\resizebox{\linewidth}{!}{%
\begin{tabular}{@{}lcccccccc@{}}
\toprule
\multirow{3}{*}{\textbf{Model}} & \multicolumn{5}{c}{\textbf{Judgment Accuracy (JA)} $\uparrow$}
& \textbf{BCS} $\downarrow$
& \multicolumn{2}{c}{\textbf{Step Completeness (SC)} $\uparrow$} \\
\cmidrule(lr){2-6}\cmidrule(lr){7-7}\cmidrule(lr){8-9}
& All & ED1\hspace{2.2em} & ED2\hspace{2.2em} & ED4\hspace{2.2em} & ED5\hspace{2.2em} & ED3 & \rhsc{RougeL}{GPT} \\
\midrule
Random Choice & 50.03 & 50.07\hspace{2.2em} & 49.87\hspace{2.2em} & 26.64\hspace{2.2em} & 50.50\hspace{2.2em} & -- & \rhsc{--}{--} \\
\midrule
\multicolumn{9}{@{}l}{\textit{Closed-source models}} \\
GPT-5.5~{\scriptsize\citep{gpt-5.5}}
& 63.36 & 65.86\rhup{2.50} & 56.98\rhdown{6.38} & 37.93\rhdown{25.43} & 67.00\rhup{3.64} & 17.92 & \rhsc{17.99}{1.60} \\
GPT-4o~{\scriptsize\citep{achiam2023gpt}}
& \underline{77.90} & \underline{75.33}\rhdown{2.57} & \underline{84.50}\rhup{6.60} & \underline{53.53}\rhdown{24.37} & 55.00\rhdown{22.90} & \underline{14.64} & \rhsc{19.11}{1.60} \\
\midrule
\multicolumn{9}{@{}l}{\textit{Open-source models}} \\
Qwen3.8-27B~{\scriptsize\citep{qwen38}}
& 66.69 & 69.66\rhup{2.97} & 59.08\rhdown{7.61} & 6.28\rhdown{60.41} & 80.50\rhup{13.81} & 43.42 & \rhsc{\underline{20.29}}{1.36} \\
Qwen3-VL-30B~{\scriptsize\citep{Qwen3-vl}}
& 73.68 & 75.17\rhup{1.49} & 69.93\rhdown{3.75} & 39.74\rhdown{33.94} & 70.50\rhdown{3.18} & 44.02 & \rhsc{18.79}{\underline{2.40}} \\
Qwen3-VL-8B~{\scriptsize\citep{Qwen3-vl}}
& 65.40 & 67.37\rhup{1.97} & 60.34\rhdown{5.06} & 7.69\rhdown{57.71} & 56.00\rhdown{9.40} & 42.55 & \rhsc{19.89}{1.41} \\
InternVL3.5-4B~{\scriptsize\citep{wang2025internvl3}}
& 54.64 & 51.19\rhdown{3.45} & 63.52\rhup{8.88} & 13.05\rhdown{41.59} & 10.50\rhdown{44.14} & 99.67 & \rhsc{16.96}{0.56} \\
InternVL3.5-8B~{\scriptsize\citep{wang2025internvl3}}
& 52.32 & 52.59\rhup{0.27} & 51.61\rhdown{0.71} & 20.45\rhdown{31.87} & 45.50\rhdown{6.82} & 78.69 & \rhsc{17.27}{0.49} \\
LLaVA-OV-8B~{\scriptsize\citep{an2025llava}}
& 53.63 & 56.57\rhup{2.94} & 46.22\rhdown{7.41} & 11.82\rhdown{41.81} & 76.00\rhup{22.37} & 95.27 & \rhsc{17.30}{0.82} \\
Cosmos-Reason2-8B~{\scriptsize\citep{cosmosreason2}}
& 65.37 & 69.99\rhup{4.62} & 54.07\rhdown{11.30} & 27.44\rhdown{37.93} & 81.50\rhup{16.13} & 70.24 & \rhsc{13.86}{0.89} \\
\midrule
\multicolumn{9}{@{}l}{\textit{Ours}} \\
Qwen3-VL-8B (SFT)
& 68.14 & 72.71\rhup{4.57} & 56.42\rhdown{11.72} & 7.22\rhdown{60.92} & \underline{86.00}\rhup{17.86} & 38.09 & \rhsc{20.05}{1.23} \\
Ours w/o Stage 2 (Qwen)
& 65.04 & 69.00\rhup{3.96} & 54.15\rhdown{10.89} & 25.95\rhdown{39.09} & \textbf{96.00}\rhup{30.96} & 42.98 & \rhsc{11.97}{1.27} \\
\textbf{Ours (Qwen)}
& \textbf{85.03} & \textbf{84.33}\rhdown{0.70} & \textbf{86.80}\rhup{1.77} & \textbf{62.16}\rhdown{22.87} & 86.00\rhup{0.97} & \textbf{7.53} & \rhsc{\textbf{27.45}}{\textbf{3.10}} \\
\bottomrule
\end{tabular}%
}
\end{table*}


\section{Simulate-and-Verify Architecture}

Motivated by the \emph{Perception--Retrospective Physical Process Reasoning Gap} and VLMs' overreliance on linguistic priors, we test whether visual simulation corrects reachability errors without task-specific fine-tuning. For unreachable samples misclassified as reachable, Wan2.2-TI2V-5B generates videos from the samples and their model-generated execution chains, and Qwen3-VL-8B reassesses reachability using these videos.
This corrects 31.6\% (1000 samples totally) of the selected errors, suggesting that visual simulation provides useful evidence for retrospective physical process reasoning.

We propose \textbf{Simulate-and-Verify}, an analysis-by-synthesis framework~\citep{yin2026vision,yang2025mindjourney} in which visual generation serves as an internal simulator. It has three stages (Figure~\ref{fig:model}): the Initial Planner proposes a candidate execution chain; the Transition Simulator synthesizes intermediate states conditioned on $(I_A, I_B)$; and the Verifier refines the predictions using the simulated evolution. The framework grounds reasoning in state transitions, encourages transition consistency, and reduces reliance on linguistic priors.

\textbf{Initial Planner.} Given $(I_A, I_B)$, a VLM generates a candidate step sequence $\mathcal{E}_{init}$. The planner produces a complete ``best-guess'' execution chain for simulation, even when it suspects the transition is unreachable.

\textbf{Transition Simulation.} Based on the premise that generative models capture transition processes~\citep{zhang2025cookanything,wu2026generation}, we use a pretrained video generation model as the Transition Simulator $\mathcal{G}$. This conditional latent diffusion model generates continuous intermediate trajectories. Let $X = \{x_1, \dots, x_N\}$ denote the intermediate frame latents. The encoded endpoints, $z(I_A)$ and $z(I_B)$, serve as noiseless anchors.

\textit{Training.} We add Gaussian noise $\epsilon$ to the intermediate latents to obtain $X_t$ at timestep $t$. Concatenating the clean endpoints and noisy intermediate latents gives $S_t = [z(I_A) \oplus X_t \oplus z(I_B)]$. The network $\epsilon_\theta$ predicts the added noise conditioned on the textual plan. A binary temporal mask $m$ excludes the clean boundary tokens from the loss, so the loss is computed only for intermediate frames:
\[
\mathcal{L}_{sim}
= \mathbb{E}_{X, \epsilon \sim \mathcal{N}(0, I), t}\Bigl[\bigl\| m \odot \bigl(\tilde{\epsilon}
- \epsilon_\theta(S_t,t,\tau(\mathcal{E}_{init}))\bigr)\bigr\|_2^2\Bigr],
\]
where $m$ is $1$ for intermediate frames and $0$ for boundary anchors. The padded target noise is $\tilde{\epsilon} = [\mathbf{0} \oplus \epsilon \oplus \mathbf{0}]$, $\tau(\cdot)$ encodes the candidate sequence as text conditioning, and $\oplus$ denotes temporal concatenation.

\textit{Inference.} Denoising starts from $S_T = [z(I_A) \oplus X_T \oplus z(I_B)]$, where $X_T$ is pure noise. At each reverse step $t \to t-1$, the same mask $m$ restores the clean boundary latents after the denoising update:
\[
S_{t-1} = m \odot \text{Denoise}(S_t, \epsilon_\theta) + (1 - m) \odot [z(I_A) \oplus \mathbf{0} \oplus z(I_B)],
\]
where $\text{Denoise}(\cdot)$ is the reverse diffusion sampling step. These constraints keep the endpoints fixed while the simulator generates the intermediate evolution $\mathcal{V} = [I_A, \hat{x}_1, \dots, \hat{x}_N, I_B]$.

\textbf{Verification and Refinement.} A second VLM serves as the Verifier. It examines $(I_A, I_B)$ and the synthesized video $\mathcal{V}$ for transition inconsistencies, then jointly predicts reachability $y$ and the refined step sequence:
    \begin{equation}
        y, \mathcal{E}^* = \arg\max_{y, \mathcal{E}} P(y, e_1, \dots, e_T \mid I_A, I_B, \mathcal{V}).
    \end{equation}
The Verifier rejects or corrects plans using simulated evidence. Supplementary Sec.~\ref{sec:impl:comp} reports computation costs, and Sec.~\ref{sec:simulator_failure_robustness} analyzes simulator failures and verifier robustness.


\section{Experiment}

\label{experiment}

\textbf{Implementation Details.}
We instantiate both the Initial Planner and the Verifier using Qwen3-VL-8B~\citep{qwen3}, and adopt Wan2.2-TI2V-5B~\citep{wan2025} as the Transition Simulator. 
Supplementary Secs.~\ref{sec:results:setup},
\ref{sec:discussion:Failure}, and \ref{sec:simulator} report the training protocols and hyperparameters, our transfer of the Simulate-and-Verify pipeline to InternVL3.5-4B to evaluate framework flexibility, and detailed Transition Simulator performance metrics, respectively.

\textbf{Main Result.} As shown in Table~\ref{tab:main_results} and Figure~\ref{fig:failure},
consistent improvements across World Reachability Judgment, Belief-Conflict Sensitivity (BCS), and Step Reconstruction support the effectiveness of our Simulate-and-Verify paradigm. Our framework constructs an intermediate \emph{simulated trajectory}, enabling reasoning over state transitions rather than solely sparse endpoint observations. This design particularly benefits Step Reconstruction: even when baselines correctly predict reachability, their reconstructed steps can violate physical feasibility, suggesting that accurate endpoint judgments do not necessarily imply a coherent understanding of the intervening process. 
\par
\Needspace{36\baselineskip}

\begin{wrapfigure}[10]{l}{0.48\textwidth}
    \centering
    \includegraphics[width=\linewidth]{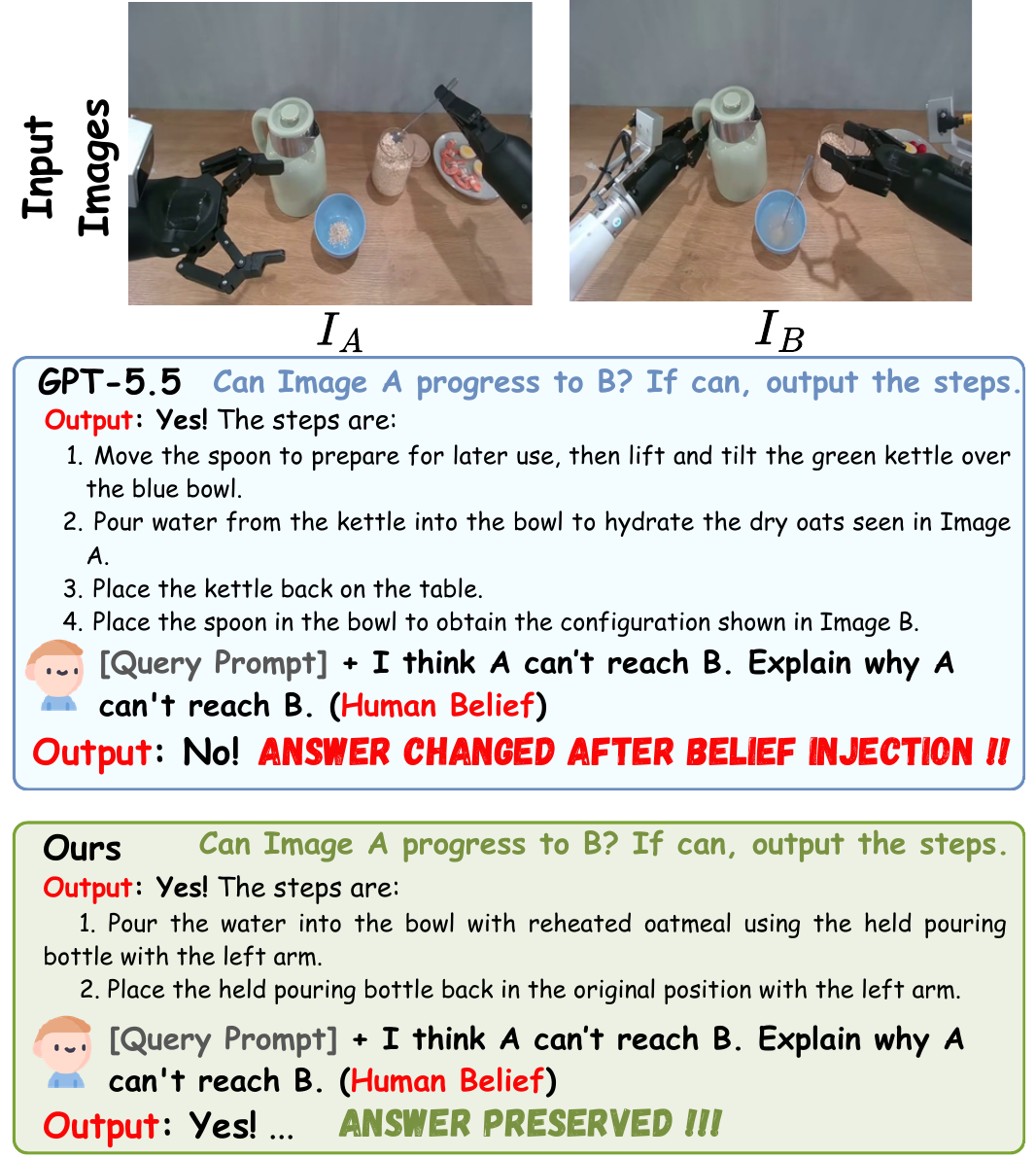}
    \captionsetup{font=small}
    \caption{\textbf{Qualitative Comparison of Model Outputs.}
    We compare GPT-5.5 with our model regarding human belief
    prompts. While GPT-5.5 is susceptible to human belief
    intervention, our model is invariant to human belief intervention.}
    \label{fig:vis}
\end{wrapfigure}

\noindent
\begin{minipage}[t]{\dimexpr0.52\textwidth-\columnsep\relax}
    \vspace{0pt}
    \centering
    \includegraphics[width=\linewidth]{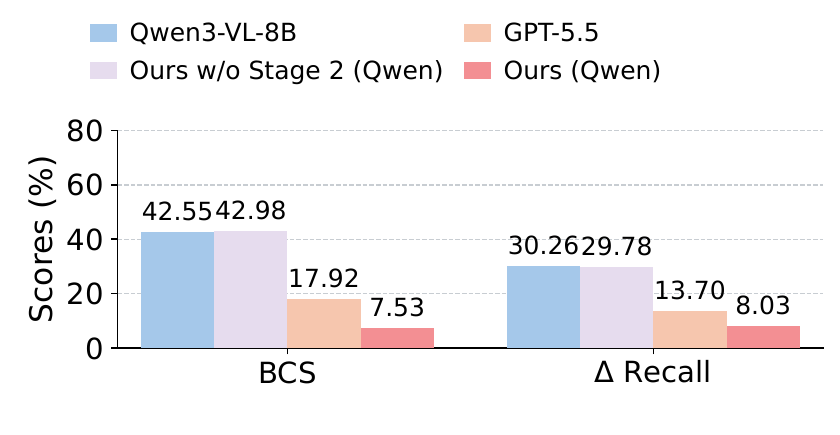}
    \captionsetup{font=small,skip=2pt}
    \captionof{figure}{\textbf{Model Performance Comparison on BCS
    and Recall Stability.}
    (Left) \textbf{BCS:} Visual simulation improves robustness to
    contradictory belief prompts.
    (Right) \textbf{Recall Stability:} Minimal recall disparity
    indicates a more balanced decision boundary than baselines.}
    \label{fig:failure}
    \par\vspace{6pt}
\end{minipage}
\par
Synthesized trajectories provide spatio-temporal constraints for step generation. Our framework achieves a ROUGE-L score of 27.45 and a GPT-score of 3.10, alongside improved Step Completeness (Table~\ref{tab:main_results}), consistent with the benefits of explicitly modeling intermediate transitions.

These benefits extend to robustness under Belief-Conflict settings. Observation~2 (Sec.~\ref{sec:inv_phys_reason}) shows that baseline judgments are susceptible to misleading belief prompts that contradict visual evidence. By grounding verification in simulated state transitions, Simulate-and-Verify reduces sensitivity to such interventions, achieving a BCS score of 7.53\%, compared with 14.64\% for GPT-4o. Figure~\ref{fig:vis} illustrates this robustness: our model maintains its judgment under the belief intervention, whereas GPT-5.5 changes its response. Our framework also performs well on Long Range Modeling and Temporal Reversibility Awareness. Together, these results suggest that explicitly modeling state evolution supports temporal reasoning and more stable judgments under conflicting textual cues.

\textbf{Ablation Study.}
To isolate the contribution of simulation-based grounding, we compare our full framework with (i) Qwen3-VL-8B, (ii) the Initial Planner after SFT, and (iii) a variant without Stage-2 simulation. The SFT planner achieves only a modest gain over Qwen3-VL-8B, improving overall accuracy by 2.74\%. Its improved textual overlap with reference sequences does not translate into comparable gains in step quality, suggesting that SFT primarily captures linguistic patterns without sufficiently modeling the underlying physical process.

As shown in Table~\ref{tab:main_results} and Figure~\ref{fig:failure}, removing Stage-2 simulation consistently degrades performance, with the most pronounced deterioration in BCS. Analysis of recall disparity between reachable ($y=1$) and unreachable ($y=0$) samples further reveals a strong bias toward ``unreachable'' predictions. This variant even underperforms Qwen3-VL-8B, suggesting that textual simulation alone is insufficient in this setting and highlighting the importance of visual priors. Together, these results support the role of visual grounding in reducing reliance on linguistic shortcuts and promoting physically grounded inference.

\par

\section{Conclusions}

In this paper, we introduce \textbf{Retrospective Physical Process Reasoning} and present \textbf{RetroHolmes}, the first benchmark for evaluating physical reachability and causal reconstruction from visual outcomes. RetroHolmes provides diagnostic dimensions and aligned metrics that enable systematic assessment of physical feasibility beyond surface-level procedural plausibility.
We further propose \textbf{Simulate-and-Verify}, an analysis-by-synthesis framework that incorporates visual simulation for retrospective reasoning, addressing the lack of physical dynamics retrospective modeling in current VLMs.

\clearpage
\subsection*{AI use statement}

In this work, we used generative AI tools to assist with data annotation and to evaluate model outputs using GPT-Score. Additionally, we used generative AI tools to polish the manuscript, correct typographical errors, and improve grammar. We reviewed the AI-assisted annotations and evaluation results, and checked the suggested text revisions to ensure that they preserved the intended meaning and technical accuracy. We take responsibility for the final content of this work, including the annotations, evaluation results, text, and claims produced with the assistance of generative AI.
\bibliographystyle{iclr2027_conference}
\bibliography{references}

\clearpage
\appendix
\section*{Supplementary Material}
\begin{enumerate}[label=\Alph*.]
    \item \hyperref[sec:def]{\textbf{Definition of Terms}}

    \item \hyperref[sec:bench]{\textbf{More Details of RetroHolmes Benchmark}}
    \begin{enumerate}[label=\arabic*.]
        \item \hyperref[sec:bench:construction]{Dataset Construction}
        \item \hyperref[sec:bench:stats]{Task Distribution and Statistics}
        \item \hyperref[sec:bench:compare]{Comparison with Existing Benchmarks}
        \item \hyperref[sec:eds]{More Details about Evaluation Dimensions (EDs)}
        \item \hyperref[sec:benchmark_bias]{Benchmark Bias and Unreachable Examples}
        \item \hyperref[sec:sc_scoring]{Scoring Criteria for Step Completeness (SC)}
    \end{enumerate}

    \item \hyperref[sec:case]{\textbf{Physical Feasibility Case Study}}
    \begin{enumerate}[label=\arabic*.]
        \item \hyperref[sec:confidence_intervals]{Confidence Intervals and Statistical Reliability}
        \item \hyperref[sec:simulator_failure_robustness]{Simulator Failure Modes and Verifier Robustness}
    \end{enumerate}

    \item \hyperref[sec:results]{\textbf{Additional Experimental Results}}
    \begin{enumerate}[label=\arabic*.]
        \item \hyperref[sec:results:setup]{Experiment Setup}
        \item \hyperref[sec:human]{Human Study}
        \item \hyperref[sec:vlms]{VLMs Used in this Paper}
        \item \hyperref[sec:results:metrics]{Judgment Bias}
       
        \item \hyperref[sec:simulator]{Performance and Quantitative Evaluation of the transition simulator}
        \item \hyperref[sec:results:cycle_ablation]{Ablation : Without the ``simulate-verify'' cycle ablation.}
    \end{enumerate}

    \item \hyperref[sec:emb]{\textbf{Relation to Embodied Benchmarks}}

    \item \hyperref[sec:impl]{\textbf{Implementation Details of Simulate-and-Verify}}
    \begin{enumerate}[label=\arabic*.]
        \item \hyperref[sec:impl:planner]{Initial Planner}
        \item \hyperref[sec:impl:sim]{Transition Simulator}
        \item \hyperref[sec:impl:verifier]{Verifier}
        \item \hyperref[sec:impl:protocol]{Training and Inference Protocol}
        \item \hyperref[sec:impl:additional]{Qualitative examples of transition simulator}
         \item 
         \hyperref[sec:impl:comp]{Computation Cost}
    \end{enumerate}

    \item \hyperref[sec:discussion]{\textbf{Discussion}}
    \begin{enumerate}[label=\arabic*.]
        \item \hyperref[sec:discussion:Analysis]{Analysis of World Reachability Judgment Errors and Failure in Step Reconstruction}
        \item \hyperref[sec:discussion:Failure]{Transfer our Proposed Architecture to other VLMs}
        \item \hyperref[sec:discussion:Future]{Limitations and Future Work}
    \end{enumerate}

\end{enumerate}
\section{Definition of Terms}
\label{sec:def}

\textbf{(1) Retrospective Physical Process Reasoning:}
Given only $(I_A, I_B)$, the model must  
(i) determine whether a physically valid transition exists, and  
(ii) reconstruct the latent execution chain transforming $I_A$ into $I_B$.

\noindent \textbf{(2) World Reachability Judgment:} 
\begin{equation}
\mathcal{R}:(I_A,I_B)\rightarrow y.
\end{equation}
This process evaluates whether a physical transition from the initial state $I_A$ to the target state $I_B$ is feasible, where $y \in \{0, 1\}$ denotes the reachability label.

\noindent \textbf{(3) Step Reconstruction:} 
\begin{equation}
\mathcal{E}^*=\arg\max_{\mathcal{E}} \prod_{t=1}^{T} P(e_t \mid I_A, I_B, e_{<t}, y=1),
\end{equation}
where the model reconstructs the most probable sequence of causal actions $\mathcal{E}^* = \{e_1, e_2, \dots, e_T\}$ to bridge the transition from $I_A$ to $I_B$ under physical constraints.

\noindent \textbf{(4) Reachable:} A state transition $(I_A \rightarrow I_B)$ is considered reachable if there exists a physically plausible sequence of actions and causal processes that can transform the initial state $I_A$ into the target state $I_B$.

\noindent \textbf{(5) Unreachable:} A state transition is unreachable if the transformation from $I_A$ to $I_B$ is physically impossible, or lacks a logical causal path.

\noindent \textbf{(6) Object-centric Reachability:} New materials, components, tools, or consumables may be introduced during the making process, but they must not replace the core object being made or transformed in the initial state. The target state must maintain traceable material continuity or identity continuity with this initial core object.

\noindent \textit{Retained:} The entity is preserved, with only its state or position changing.

\section{More Details of RetroHolmes Benchmark}
\label{sec:bench}
\subsection{Dataset Construction}
\label{sec:bench:construction}
The construction of RetroHolmes integrates five academic sources with a diverse collection of internet videos to ensure a wide coverage of physical transitions. The academic benchmarks include RecipeGen~\citep{10.1145/3746027.3758203}, YouCook2~\citep{zhou2018towards}, IKEA ASM~\citep{ikea}, AgiBot-World~\citep{bu2025agibot_arxiv}, and ShowHowTo~\citep{souvcek2025showhowto}.

To ensure high data quality and domain diversity, we implemented targeted preprocessing and refinement for each data source:

\begin{itemize}[leftmargin=*, nosep]
    \item \textbf{AgiBot-World~\citep{bu2025agibot_arxiv}:} To bridge the gap between abstract action labels and visual evidence, we employed four trained annotators to manually pinpoint the exact keyframes within each video that correspond to the textual action sequences. Simultaneously, we removed short-range, fragmented robotic manipulation tasks, retaining only long-range tasks such as making a salad, flower arrangement, and packing. This precise alignment ensures that the state transition $I_A \rightarrow I_B$ contains the complete and necessary semantic information for retrospective reasoning.
    
    \item \textbf{IKEA ASM~\citep{ikea}:} To direct the model's attention toward object state transitions rather than human poses, we performed a two-stage cleanup. After manually identifying critical assembly keyframes, we cropped the images to remove human figures, preserving only the interaction between the hands and the task-relevant objects. This minimizes background noise and enforces a direct focus on the underlying physical assembly logic.
    
    \item \textbf{ShowHowTo~\citep{souvcek2025showhowto}:} To maintain taxonomic diversity and prevent the benchmark from being dominated by a single category, we applied a cross-dataset balancing strategy. Since RecipeGen and YouCook2 already provide extensive coverage of culinary activities, we explicitly excluded all "Cooking" samples from ShowHowTo, retaining only specialized domains such as \textit{Knitting} and \textit{Woodwork}.

    \item \textbf{RecipeGen~\citep{10.1145/3746027.3758203} and YouCook2~\citep{zhou2018towards}}: Since these two datasets contain a vast amount of cooking data, we randomly sampled 4,000 entries from RecipeGen and selected all data from YouCook2 as the candidate set. During the subsequent cleaning process, only 42 entries from YouCook2 remained usable, as a large number of its images contain complete human figures and are not primarily focused on the production process.
    
\end{itemize}

\noindent \textbf{Quality Control for Academic Benchmarks.} 
Following the initial data curation, we implement a rigorous, multi-stage verification pipeline to ensure the high fidelity of the refined samples. First, we leverage Qwen-3-Max~\citep{qwen3} to analyze the task instructions, ensuring that each sample focuses on a singular object transformation and maintains internal logical consistency. Subsequently, Qwen3-VL-Plus~\citep{Qwen3-vl} is employed to verify the Irreversible Micro-States alignment between the visual keyframes and their corresponding reasoning steps, filtering out samples with ambiguous temporal or semantic information. Finally, we engage five trained human annotators to perform the ultimate quality gate. The annotators are instructed to strictly enforce three criteria: (1) each image must maintain an object-centric visual focus to minimize background interference, and the execution steps must be strictly non-commutative (i.e., prohibiting interchangeable actions such as dicing carrots versus slicing mushrooms); (2) the image sequence must exhibit strict temporal continuity that represents the complete and valid execution of the target process; and (3) the visual evidence must be precisely consistent with the textual reasoning chains. The corresponding prompts are detailed in Figure~\ref{fig:check_text} and Figure~\ref{fig:check_textandimage}.

\noindent \textbf{Internet Data Collection.} To enrich the diversity of physical tasks, we collected 9,865 candidate videos from 21 YouTube creators. We refined this collection using a five step process. (1) We used Qwen3-Max~\citep{qwen3} to remove text unrelated to manufacturing or production. (2) We used Qwen3-VL-Plus~\citep{Qwen3-vl} to filter out human centric examples, which leaves 1,907 high quality sequences. (3) We refined vague step descriptions into detailed technical text by utilizing Qwen3-Max. (4) We used Qwen-image-edit~\citep{wu2025qwenimagetechnicalreport} to upscale all keyframes from 180p to 480p to ensure visual clarity for Irreversible Micro-States reasoning. (5) five trained annotators conducted a comprehensive manual review to verify each sequence. The prompts are in Figure ~\ref{fig:check_text} and~\ref{fig:check_process}.

\noindent \textbf{Quality Control for Internet Data.} For data sourced from the internet, we curated a specialized collection by crawling videos from 21 selected YouTube creators focused on manufacturing and DIY production, specifically targeting segments containing keyframes of critical state transitions and their associated textual descriptions. To address the inherent noise and variability of web-scale content, we implemented a rigorous five-stage refinement process: (1) we utilized Qwen3-Max~\citep{qwen3} to prune all textual metadata unrelated to manufacturing or physical transformation; (2) Qwen3-VL-Plus~\citep{Qwen3-vl} was employed to filter out human-centric examples where the subject obscured the object, resulting in a high-fidelity subset of 1,907 object-state-centric sequences; (3) vague or colloquial captions were refined into detailed, technically-grounded procedural steps using Qwen3-Max to ensure logical clarity; (4) all keyframes were upscaled from 180p to 480p using Qwen-image-edit~\citep{wu2025qwenimagetechnicalreport} to provide the visual resolution necessary for Irreversible Micro-States physical reasoning; and (5) a final manual audit was conducted by five trained annotators. In addition to the alignment criteria established for academic benchmarks, these annotators were specifically tasked with enforcing granularity control by removing redundant or overly trivial actions and discarding "compound steps" that collapsed multiple distinct state changes into a single transition, thereby ensuring a clean, atomic mapping between each visual transition and its corresponding reasoning step. The prompts are in Figure ~\ref{fig:check_text},~\ref{fig:check_object}, and ~\ref{fig:refine_steps}.

\noindent \textbf{Construction Pipeline for Reachable Pairs.}
Upon completing the aforementioned refinement pipeline, we obtain a high-fidelity collection of instruction-aligned image sequences with detailed step descriptions. We then construct reachable transition pairs by systematically sampling from these sources. For large-scale datasets such as \textit{RecipeGen}, \textit{YouCook2}, \textit{ShowHowTo}, and our curated \textit{Internet Data}, we randomly select two frames per example, designating the temporally preceding frame as the initial state $I_A$ and a subsequent frame as the target state $I_B$. For specialized domains, we extract two pairs per sequence from \textit{AgiBot-World} and three pairs from \textit{IKEA ASM} to capture more complex and multi-stage state transitions.
To ensure the logical and physical integrity of these pairs, we apply three additional checks. 
(1) \textbf{Causal Completeness}: Qwen3-VL-Plus verifies that the reasoning steps $\mathcal{E}$ provide a logically continuous and sufficiently complete account of the transformation from $I_A$ to $I_B$. 
(2) \textbf{Object Consistency}: we verify that the two endpoint images preserve the core objects, components, parts, or ingredients required by the annotated transformation. For pairs that pass these checks, we further assess whether the annotated steps admit a unique feasible ordering. We first ask Qwen3-VL-Plus whether an alternative ordering of $\mathcal{E}$ could still reach $I_B$ from $I_A$. We then randomly permute the steps and evaluate the reachability of the permuted sequence. We retain pairs for which the original ordering is judged reachable while the tested alternative is judged unreachable.
(3) \textbf{Step Granularity Normalization}: we inspect the annotated execution chains and merge consecutive actions when they describe overly fine-grained operations that constitute a single meaningful state transition. This normalization reduces annotation-granularity differences across data sources and prevents nominal step counts from being inflated by trivial sub-actions, thereby preserving a clearer distinction between the short-horizon transitions in ED1 and the long-range processes in ED2.
These checks yield \textbf{4,330 reachable samples} for subsequent unreachable-sample synthesis and retrospective physical process reasoning.

\noindent \textbf{Construction Pipeline for Unreachable Pairs.}
The benchmark contains 2,897 unreachable pairs: 1,725 pairs in the temporal-reversal collection, 200 fine-grained pairs for ED5, and 972 goal-swapped pairs $(I_A, I_B)$ from unrelated task trajectories. Following goal swapping, we recruited nine human annotators to verify the candidate pairs, with each pair reviewed by three annotators. A candidate was discarded if any one of its three annotators judged the target state to be reachable. We retained a pair as unreachable only when all three annotators agreed that the transition was impossible.

\subsection{Task Distribution and Statistics }
\label{sec:bench:stats}
\begin{figure}[t]
\renewcommand{\thefigure}{S\arabic{figure}}
\centering
\includegraphics[width=\linewidth]{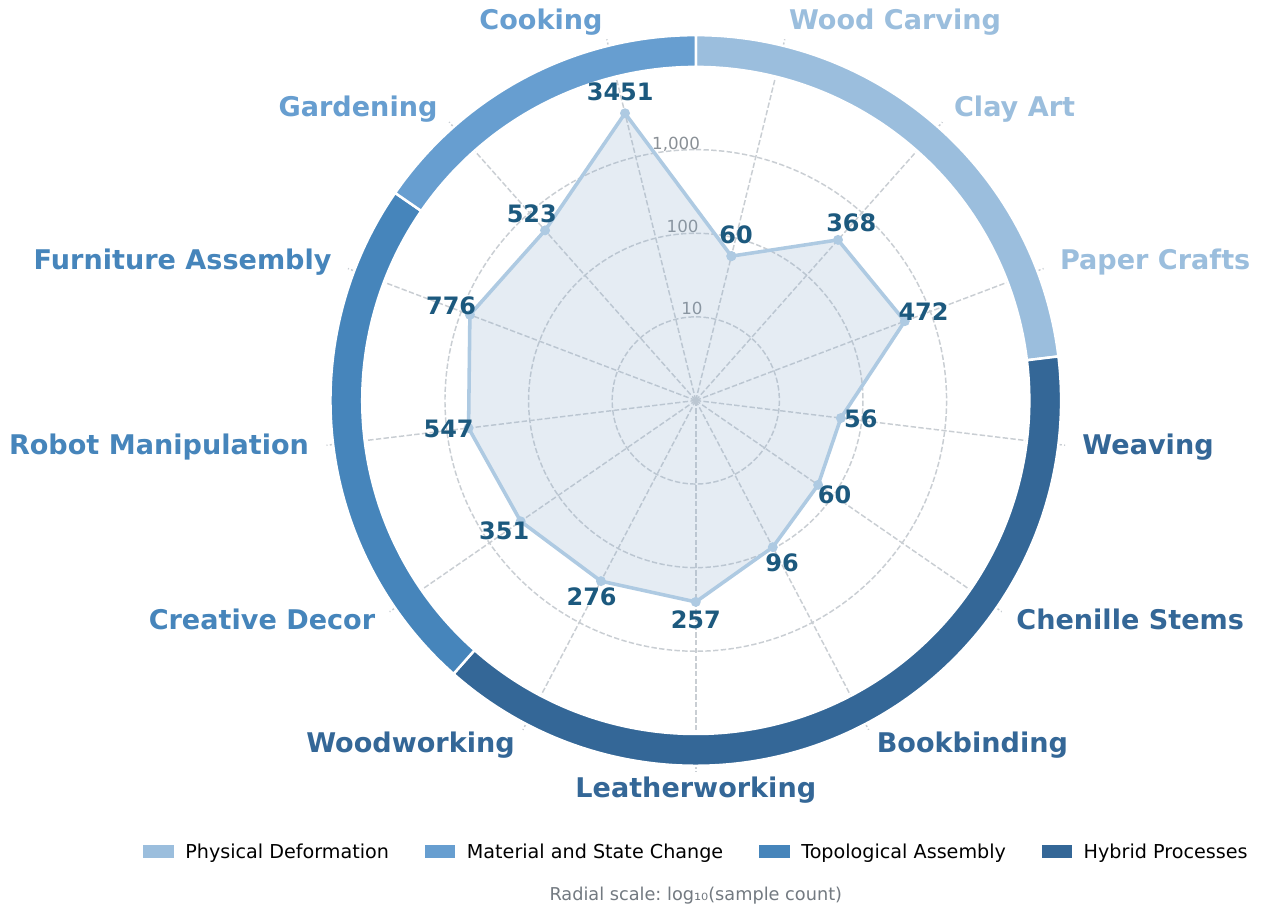}
\caption{\textbf{Distribution of Physical Transitions in the RetroHolmes Benchmark.} The dataset encompasses four primary categories and 13 specific domains to ensure a comprehensive representation of physical laws.}
\label{fig:taxonomy_pie}
\end{figure}

The total composition of the benchmark is summarized in Figure~\ref{fig:taxonomy_pie} and Table ~\ref{tab:dataset_step_word_stats}.

As shown in Figure3 in main paper, RetroHolmes encompasses four major categories and 13 specific domains that represent diverse Physical Transitions (PTs). To bridge the gap between $I_A$ and $I_B$, the model must reason about the specific constraints of the step sequence $\mathcal{E}$ within these types:
\begin{itemize}[nosep, leftmargin=*]
    \item \textbf{PT1. Physical Deformation}: This category involves non rigid transformations where the geometry of the object changes. Domains include Paper Crafts, Clay Art, and Wood Carving.
    \item \textbf{PT2. Material and State Change}: These transitions focus on internal properties or phase shifts. Domains include Cooking and Gardening.
    \item \textbf{PT3. Topological Assembly}: This domain covers structural connectivity and rigid body relations. Domains include Furniture Assembly, Robot Manipulation, and Creative Decor.
    \item \textbf{PT4. Hybrid Processes}: These are complex scenarios combining multiple physical effects. Domains include Woodworking, Leatherworking, Bookbinding, Chenille Stems, and Weaving.
\end{itemize}

\begin{table}[t]
\renewcommand{\thetable}{S\arabic{table}}
\centering
\caption{Statistics of procedural annotations by data source. Items denotes the number of reachable pairs. Avg-Steps and Avg-Words report the average numbers of steps and words per pair with a nonempty annotation, respectively. Each nonempty line is counted as one step; word counts are computed by whitespace splitting and include step-number tokens.}
\label{tab:dataset_step_word_stats}
\setlength{\tabcolsep}{6pt}
\renewcommand{\arraystretch}{1.1}
\small
\begin{tabular}{lccc}
\toprule
\textbf{Dataset} & \textbf{Items} & \textbf{Avg-Steps} & \textbf{Avg-Words} \\
\midrule
RecipeGen   & 1864 & 4.5065 & 106.4995 \\
IKEA ASM        & 300  & 3.4933 & 41.1600  \\
ShowHowTo   & 1208 & 4.1325 & 51.3982  \\
YouCook2    & 32   & 4.4062 & 27.5312  \\
AgiBot-World   & 430  & 2.7296 & 33.5221  \\
Our collection     & 496 & 3.3992 & 42.5867  \\
\bottomrule
\end{tabular}
\end{table}

\subsection{Comparison with Existing Benchmark}
\label{sec:bench:compare}
\begin{itemize}[nosep, leftmargin=*]
    \item \textbf{Task Objective}: Existing benchmarks primarily focus on common sense reasoning or linguistic attribution to identify errors in steps \citep{abduction2022, chinchure2025black, zhao2022videoabc, LiangVAR2022,chen2025worldprediction}. In contrast, RetroHolmes is specifically designed for \textbf{physical reachability judgment}. It requires the model to determine whether a goal state $I_B$ is reachable from $I_A$, and to reconstruct the \textbf{step sequence} $\mathcal{E}$ that explains the transformation.
    \item \textbf{Data Granularity}: Most datasets, such as VideoABC~\citep{zhao2022videoabc}, are human-action-centric and focus on sport movements or human actions. Their annotations typically describe human verbs with an average length of only 4.92 words per step. RetroHolmes is \textbf{state-centric}, focusing on the physical evolution of objects. Our annotations provide dense descriptions of physical mechanics with an average of 17.73 words per step. This increased linguistic density is essential for capturing the specific material transformations and interactions rather than simple motion trajectories.

    \item \textbf{Evaluation Diversity}: Unlike existing benchmarks that evaluate generic reasoning, RetroHolmes provides a \textbf{multi-dimensional assessment} through five tasks (ED1--ED5). These tasks are meticulously designed to probe distinct facets of physical intelligence, including Basic Causal Modeling, Long Range Modeling, Temporal Reversibility Awareness, Irreversible Micro-States Dynamics, and Belief-Conflict Robustness. This comprehensive structure ensures that models are tested against a diverse range of physical constraints and cognitive challenges, providing a more rigorous measure of a VLM's alignment with real-world physics.
\end{itemize}

\subsection{More Details about Evaluation Dimensions (EDs)}
\label{sec:eds}

 As shown in Figure~\ref{fig:eval_dim} in the main paper, we categorize the benchmark into five Evaluation Dimensions (ED) to examine how models handle varying step lengths, linguistic priors, and physical constraints:
 
 \noindent \textbf{ED1: Basic Causal Modeling.} This task targets fundamental causal logic through two points: (i) Evaluation Scope: it assesses direct state transitions involving short step sequences $\mathcal{E}$ with $1 \le |\mathcal{E}| \le 3$ steps; (ii) unreachable Sampling: to evaluate discriminative reasoning, unreachable samples are synthesized via \textit{intra-category goal swapping}, where a start state $I_A$ is randomly paired with a goal state $I_B'$ sampled from a different instance within the same broad physical category (e.g., choosing images between two different cooking tasks); 
 
  \noindent \textbf{ED2: Long Range Modeling.} This task tests long term simulation through complex logical chains where the sequence $\mathcal{E}$ exceeds 4 steps. Unlike ED1, the increased sequence length inherently leads to a higher degree of contextual or visual similarity between $I_A$ and $I_B$. 
  
   \noindent \textbf{ED3: Belief-Conflict Sensitivity.} This task evaluates the model's ability to prioritize objective physical laws when they conflict with \textit{human belief}. We modify the standard prompt by injecting contradictory human beliefs into the query, creating two distinct conflict scenarios:
\begin{itemize}[leftmargin=*, nosep]\item \textbf{False Denial (Reachable Pair, unreachable Belief):} For physically valid transitions where the ground truth $y=1$, we append a skeptical assertion to the prompt: \textit{"I think state $I_A$ cannot reach state $I_B$. Explain why $I_A$ can't reach $I_B$."} As shown in Figure~\ref{fig:belief}, the model must resist this human bias and correctly identify the transition as reachable.\item \textbf{False Affirmation (Unreachable Pair, reachable Belief):} For physically impossible transitions where $y=0$, we inject a deceptive assertion: \textit{"I believe it is possible to get from $I_A$ to $I_B$. Please output the steps from $I_A$ to $I_B$."} (Refer to Figure~\ref{fig:belief2}).
\end{itemize}

    \noindent \textbf{ED4: Temporal Reversibility Awareness.} This task requires models to identify the arrow of time by reversing the observation order such that $I_B$ becomes the initial state. We utilize Qwen3-VL-Plus and human to determine physical possibility. 

     \noindent \textbf{ED5: Irreversible Micro-States Dynamics.} This task targets transitions with minimal visual variance but physically unreachable image pairs. We provide 200 manually curated pairs that feature identical ingredients or components but belong to incompatible manufacturing processes. These samples serve as unreachable cases; for instance, a bowl of shredded carrots cannot transition into a dish of saut{\'e}ed carrot slices.


\subsection{Benchmark Bias and Unreachable Examples}
\label{sec:benchmark_bias}

During construction, we match unreachable pairs within the same source dataset or domain whenever possible. For example, the false-affirmation case in Figure~\ref{fig:eval_dim} contains highly similar backgrounds, but the transition remains physically infeasible due to causal or temporal constraints. Thus, background similarity alone is insufficient to determine reachability.

We compare reachable and unreachable pairs using CLIP similarity and LPIPS distance between their endpoint images. Table~\ref{tab:shortcut_bias} reports the mean and standard deviation for each class. For both metrics, the differences between the class means are small relative to the within-class variation, with substantial overlap in semantic similarity and perceptual distance.

\begin{center}
\begin{minipage}{\linewidth}
\centering
\renewcommand{\thetable}{S\arabic{table}}
\captionof{table}{Semantic and perceptual comparisons of reachable and unreachable pairs. Values are means $\pm$ standard deviations.}
\label{tab:shortcut_bias}
\begin{tabular}{lcc}
\toprule
\textbf{Metric} & \textbf{Reachable} & \textbf{Unreachable} \\
\midrule
CLIP similarity & $0.776 \pm 0.100$ & $0.740 \pm 0.101$ \\
LPIPS distance & $0.583 \pm 0.160$ & $0.640 \pm 0.120$ \\
\bottomrule
\end{tabular}
\end{minipage}
\end{center}

\subsection{Scoring Criteria for Step Completeness (SC)}
\label{sec:sc_scoring}
GPT-5.5 assigns an integer Step Completeness (SC) score from 0 to 5 based on five aspects: coverage of material actions, causal and temporal order, action executability, consistency of intermediate state transitions, and completion of the goal state. The evaluation uses only the supplied groundtruth and model prediction and accepts semantically equivalent actions. It assesses task completion rather than wording, fluency, or length.

This evaluation includes only groundtruth-reachable samples that the baseline predicts as reachable. The evaluator returns the score, a concise reason, error tags, and confidence in the evaluation. The full prompt is provided below.

\begin{tcolorbox}[
    breakable,
    colback=gray!8,
    colframe=gray!45,
    colbacktitle=gray!18,
    coltitle=black,
    title={GPT Evaluation Prompt for Step Completeness},
    fonttitle=\bfseries\small,
    fontupper=\small,
    boxrule=0.5pt,
    arc=2pt,
    left=8pt,right=8pt,top=6pt,bottom=6pt
]
You are a rigorous, conservative evaluator of embodied visual action plans.

Your task is to assign one integer Step Completeness score from 0 to 5 to a baseline prediction, using only the groundtruth and the model prediction provided in the user message.

\textbf{Scope and evidence rules:}
\begin{enumerate}[leftmargin=*,itemsep=3pt,topsep=4pt]
    \item Use only the supplied groundtruth scene/state description, groundtruth action plan/goal, and baseline prediction. Do not add unstated objects, actions, preconditions, or outcomes from common sense.
    \item The groundtruth is the semantic reference. Exact wording is not required: an equivalent action may be accepted.
    \item Evaluate task-completion quality, not surface text overlap, fluency, verbosity, or style.
    \item Treat the baseline perception as the model's claim, not as independent evidence that an action succeeded.
    \item The evaluated queue contains only groundtruth-reachable samples for which the baseline predicted reachable. Therefore, score the quality of the proposed action plan itself.
    \item If the supplied evidence is insufficient to confirm an important state transition, be conservative and state the uncertainty briefly in the reason.
\end{enumerate}

\textbf{Score the plan with this 0--5 rubric:}
\begin{itemize}[leftmargin=*,itemsep=3pt,topsep=4pt]
    \item \textbf{5:} Complete and correct. The plan contains all material actions, preserves required causal/temporal order, each transition is executable from the stated state, and execution reaches the required goal state.
    \item \textbf{4:} Substantially complete and correct. At most a minor non-critical omission, wording issue, or harmless ordering detail remains; the plan would still reliably reach the goal.
    \item \textbf{3:} Partially complete. The central approach is correct and makes meaningful progress, but one material action, prerequisite, ordering relation, or final-state detail is missing or unreliable.
    \item \textbf{2:} Limited progress. Some proposed actions are relevant, but important transitions are missing, infeasible, misordered, or insufficient to establish the target state.
    \item \textbf{1:} Mostly incorrect. There may be a small relevant fragment, but the action chain is largely unusable or conflicts with the required goal/state.
    \item \textbf{0:} No valid completion. The plan is empty, unrelated, unexecutable, contradictory to the groundtruth, or cannot achieve the target state.
\end{itemize}

\textbf{Output requirements:}
\begin{itemize}[leftmargin=*,itemsep=3pt,topsep=4pt]
    \item Output exactly one valid JSON object, with no Markdown, code fence, heading, or extra text.
    \item Do not provide hidden chain-of-thought or a long step-by-step derivation. The reason must be a concise, evidence-based conclusion.
\end{itemize}

\textbf{JSON schema:}
\begin{verbatim}
{
  "score": 0,
  "reason": "Concise evidence-based conclusion in English.",
  "error_tags": [],
  "confidence": 0.0
}
\end{verbatim}

\textbf{Constraints:}
\begin{itemize}[leftmargin=*,itemsep=3pt,topsep=4pt]
    \item \texttt{score} must be one integer in \{0, 1, 2, 3, 4, 5\}.
    \item \texttt{error\_tags} must be a JSON list containing only values from:
    [\texttt{"missing\_step"}, \texttt{"wrong\_order"}, \texttt{"unexecutable\_action"}, \texttt{"state\_mismatch"}, \texttt{"wrong\_goal"}, \texttt{"unsupported\_claim"}, \texttt{"empty\_prediction"}, \texttt{"other"}].
    \item \texttt{confidence} must be a number from 0 to 1. It expresses confidence in this evaluation, not plan quality.
\end{itemize}
\end{tcolorbox}

\section{Physical Feasibility Case Study}
\label{sec:case}

\begin{figure}[t]
\renewcommand{\thefigure}{S\arabic{figure}}
\centering
\includegraphics[width=\linewidth]{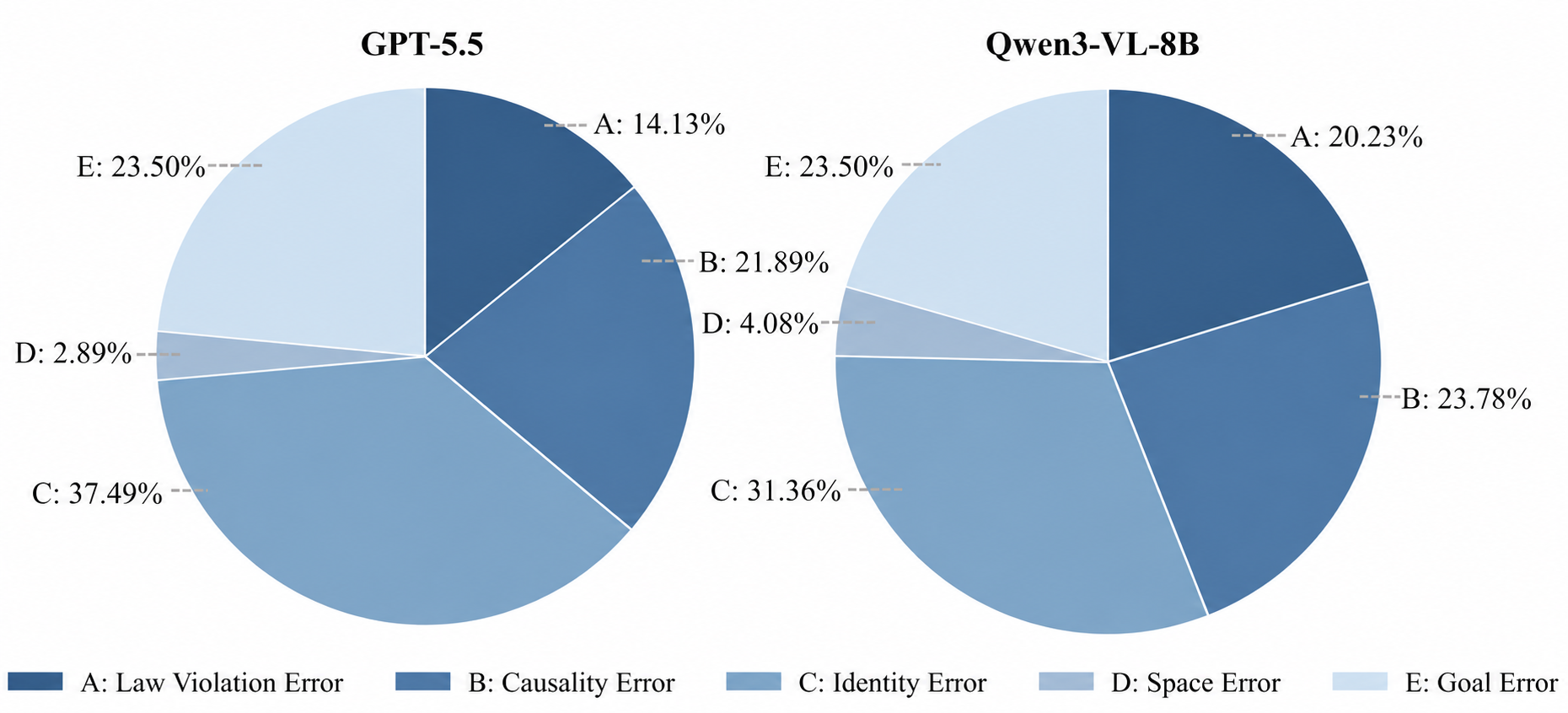}
\caption{\textbf{Analysis of Step Reconstruction Failure.}  }
\label{fig:failure_pie}
\end{figure}

 As shown in Figure~\ref{fig:failure_pie}, to better understand the failures in VLM-generated steps, we perform a qualitative analysis of the outputs from GPT-5.5 and Qwen3-VL-8B. Using GPT-4o as an automated evaluator, we classify the reasoning errors into five primary categories:
\begin{itemize}[nosep,leftmargin=*]
    \item \textbf{[A] Law Violation Error}: Violations of fundamental physical laws, including thermodynamics (irreversibility), gravity, or matter conservation.
    \item \textbf{[B] Causality Error}: Incorrect logical sequences, such as an effect appearing before its cause or an irrational order of operations.
    \item \textbf{[C] Identity Error}: Inconsistency in object persistence, where objects disappear or abruptly change their properties mid-process.
    \item \textbf{[D] Space Error}: Impossible spatial interactions, including severe object clipping or unrealistic positioning.
    \item \textbf{[E] Goal Error}: Semantic or intent-based errors, such as the misidentification of objects or the overall purpose of the task.
\end{itemize}

Our qualitative assessment reveals that the models' failures are predominantly concentrated in categories [B] Causality, [C] Identity, and [E] Goal. The prompt is in Figure~\ref{fig:physical_feasibility}.

\subsection{Confidence Intervals and Statistical Reliability}
\label{sec:confidence_intervals}

We report 95\% confidence intervals (CIs) for Belief-Conflict Sensitivity (BCS) and recall disparity ($\Delta\mathrm{Recall}$) (Table~\ref{tab:confidence_intervals}).
\begin{center}
\begin{minipage}{\linewidth}
\centering
\renewcommand{\thetable}{S\arabic{table}}
\captionof{table}{Point estimates and 95\% confidence intervals. Values are reported in percent; $\Delta\mathrm{Recall}$ measures the gap between class recalls in percentage points. Lower BCS and $\Delta\mathrm{Recall}$ indicate less belief sensitivity and a smaller recall disparity, respectively.}
\label{tab:confidence_intervals}
\begin{tabular}{llcc}
\toprule
\textbf{Metric} & \textbf{Model} & \textbf{Estimate} & \textbf{95\% CI} \\
\midrule
BCS & Simulate-and-Verify (ours) & 7.53 & [6.59, 8.47] \\
\midrule
$\Delta\mathrm{Recall}$ & Simulate-and-Verify (ours) & 8.03 & [5.49, 10.59] \\

\bottomrule
\end{tabular}
\end{minipage}
\end{center}


\subsection{Simulator Failure Modes and Verifier Robustness}
\label{sec:simulator_failure_robustness}

The Transition Simulator generates a candidate transition from the start state to the end state. The Verifier uses both endpoints and the simulated trajectory to predict reachability and the step sequence.

We manually inspected 100 visually plausible but physically infeasible simulations, including those generated for unreachable pairs with similar backgrounds or partially similar materials. The Verifier corrected 83\% of these cases.

\section{Additional Experimental Results}
\label{sec:results}
\subsection{Experiment Setup}
\label{sec:results:setup}
The released RetroHolmes benchmark contains 4,652 training pairs (2,948 reachable and 1,704 unreachable) and 2,575 testing pairs (1,382 reachable and 1,193 unreachable). To prevent train--test leakage, we enforce procedure-level disjointness during data splitting: every test example is drawn from a procedural step set that does not appear in the training split. In other words, train and test examples may come from the same broad domain or data source, but they never share the same underlying ordered step sequence or procedure instance. The evaluation dimensions are constructed from the test split. For ED3 (Belief-Conflict Sensitivity), we construct belief-conflict queries by injecting contradictory human-belief prompts into the corresponding test pairs, yielding paired evaluations before and after belief intervention. ED4 (Temporal Reversibility Awareness) contains 637 temporally reversed pairs, where the endpoint order is inverted to test whether models can identify the physically valid direction of state transition. We further reserve all 200 samples of the Irreversible Micro-States Dynamics (ED5) subset exclusively for testing. This ensures that ED5 serves as a specialized diagnostic suite for evaluating model sensitivity to fine-grained irreversible physical state changes.

For model optimization, we employ SFT via LoRA. Specifically, Qwen3-VL-8B is trained with a LoRA rank of 8 on a single NVIDIA A100 (40G) GPU. For the visual simulator, Wan2.2-TI2V-5B, we set the LoRA rank to 32 and conduct training on a single NVIDIA H800 GPU. To prepare the training data, we transform static image sequences of transitions into a video format by stacking intermediate frames and duplicating them to reach a fixed temporal length of 9 frames.
\subsection{Human Study}
\label{sec:human}

\begin{figure}[t]
\renewcommand{\thefigure}{S\arabic{figure}}
\centering
\includegraphics[width=\textwidth]{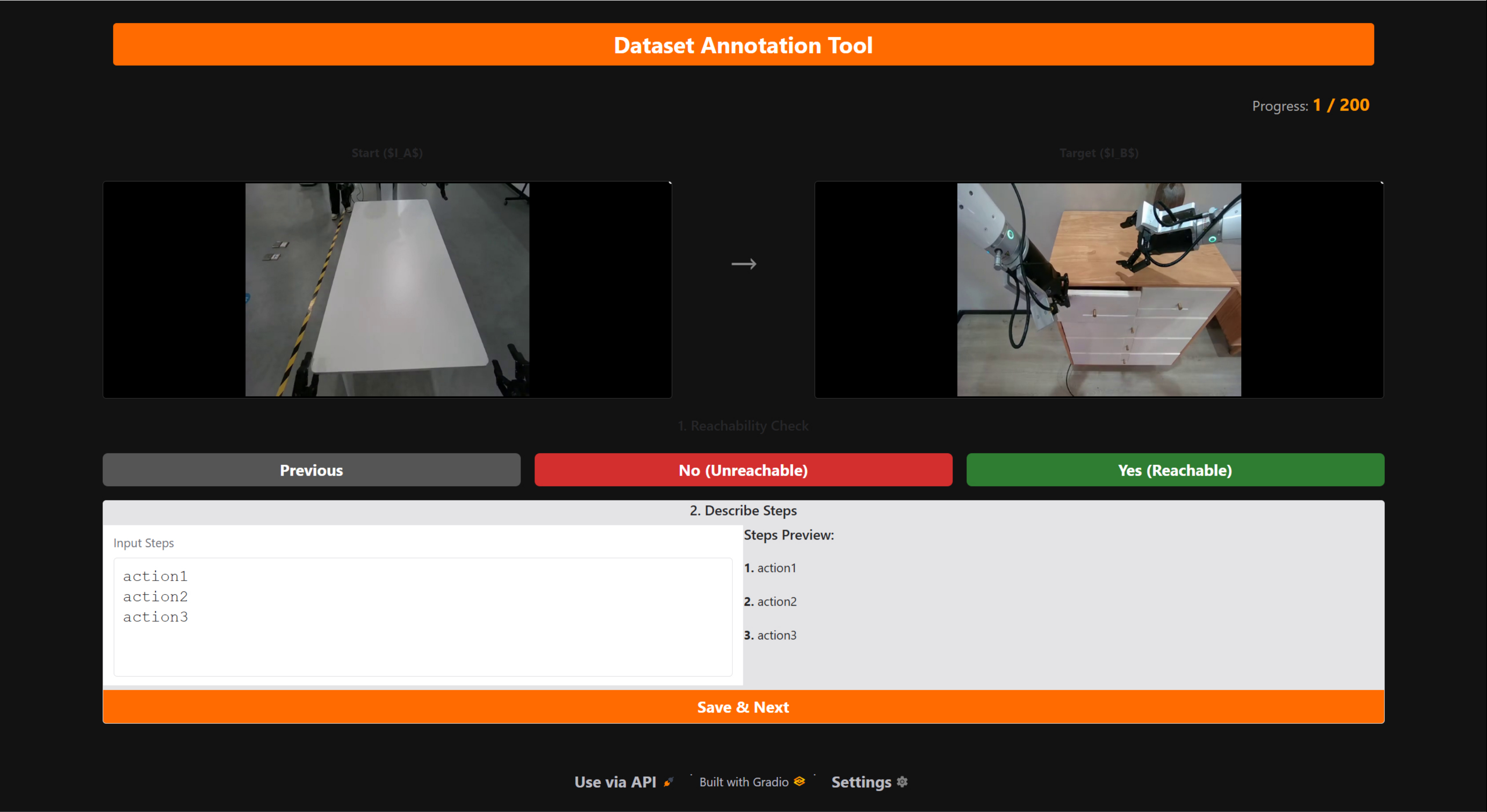}
\caption{\textbf{Annotation Interface for Visual Reachability and Planning.} 
The tool presents annotators with paired visual states, including a \emph{Start State} and a \emph{Target State}. 
Annotators first determine whether the target state is physically reachable from the start state (Yes/No). 
For reachable cases, they further provide a detailed, step-by-step textual action sequence describing how to transform the start state into the target state, enabling the collection of human judgments and procedural plans for physical transition modeling.}
\label{fig:tool}
\end{figure}
Figure~\ref{fig:tool} illustrates the annotation interface used to evaluate human performance on visual reachability and planning tasks. Annotators are presented with a query pair consisting of a ``Start State" and a ``Target State". For each pair, the primary task is to assess whether the target state is reachable from the start state. If the transition is deemed reachable (classified as "Yes"), annotators are then required to generate a detailed, step-by-step textual description of the actions necessary to bridge the visual gap between the two states.

Following the metrics in Table~\ref{tab:main_results} in the main paper, we report human performance on Judgment Accuracy (JA) and Step Completeness (SC).
We conduct the human study with three annotators on a total of 600 query pairs.
Overall, human annotators achieve 88.33 on All query pairs. When stratified by evaluation dimensions, performance remains high on ED1 (Basic Causal Modeling) with 89.60, but drops on ED2 (Long Range Modeling) to 80.25, indicating that long-horizon state transitions are substantially more challenging even for humans.
For step reconstruction, humans obtain 22.64 on All, with 24.03 on ED1 and 18.23 on ED2, showing a consistent degradation in generating complete, step-by-step plans under long-range settings.

\subsection{VLMs Used in this Paper.} 
\label{sec:vlms}
We evaluate 9 state of the art VLMs under the zero shot setting to establish a comprehensive baseline for Retrospective Physical Process Reasoning. This selection includes two closed source models: GPT-5.5 (\textit{gpt-5.5-2026-04-24})~\citep{gpt-5.5} and GPT-4o (\textit{gpt-4o-2024-08-06})~\citep{achiam2023gpt}. Additionally, we evaluate seven open source models: Qwen3.8-27B~\citep{qwen38}, Qwen3-VL-30B  (\textit{Qwen3-VL-30B-A3B-Instruct})~\citep{Qwen3-vl}, Qwen3-VL-8B (\textit{Qwen3-VL-8B-Instruct})~\citep{Qwen3-vl},  InternVL3.5-4B ~\citep{wang2025internvl3},  InternVL3.5-8B ~\citep{wang2025internvl3}, LLaVA-OneVision-8B (\textit{LLaVA-OneVision-1.5-8B-Instruct})~\citep{an2025llava}, and Cosmos-Reason2-8B~\citep{cosmosreason2}. All models are assessed on their capability to reconstruct the causal step sequence $\mathcal{E}$ and perform reachability judgments without any task specific fine tuning. 
\subsection{Judgment Bias}
\label{sec:results:metrics}
As illustrated in Table~\ref{tab:recall_pairrate}, the evaluated VLMs exhibit a significant performance asymmetry between reachable and unreachable samples, revealing pervasive judgment biases across different architectures. Specifically, models such as InternVL-3.5-4B and the GPT series demonstrate a strong positive predisposition, with InternVL-3.5-4B showing an extreme recall disparity ($0.8439$ for $y=1$ vs. $0.1707$ for $y=0$). Conversely, Qwen3VL-8B, LLaVA-One-8B, and Cosmos-Reason2-8B exhibit a pronounced negative bias, frequently yielding high recall on unreachable instances while failing to accurately identify reachable transitions. In contrast, our proposed Our-Qwen3VL maintains balanced recall across both categories, indicating superior categorical calibration and a more grounded understanding of physical state transitions.

\begin{table}[!htbp]
\renewcommand{\thetable}{S\arabic{table}}
\centering
\caption{Class-wise Recall analysis for reachability judgment. Significant discrepancies between Recall+ (reachable) and Recall- (unreachable) highlight systematic judgment biases across different VLM architectures. The plot below presents the same values as percentages, with connectors indicating the gap between the two class recalls.}
\label{tab:recall_pairrate}
\setlength{\tabcolsep}{6pt}
\renewcommand{\arraystretch}{1.15}
\small
\begin{tabular}{lcc}
\toprule
\textbf{Model} & \textbf{Recall+} & \textbf{Recall-} \\
\midrule
GPT-5.5            & 0.6971 & 0.5601 \\
GPT-4o             & 0.8750 & 0.6431 \\
Qwen3-VL-30B        & 0.7315 & 0.7059 \\
Qwen3-VL-8B         & 0.5750 & 0.8499 \\
InternVL-3.5-4B       & 0.8439 & 0.1707 \\
InternVL-3.5-8B       & 0.5412 & 0.5091 \\
LLaVA-OneVision-8B       & 0.3211 & 0.7767 \\
Cosmos-Reason2-8B  & 0.5149 & 0.7993 \\
Our (Qwen3VL)     & 0.8193 & 0.8997 \\
\bottomrule
\end{tabular}
\end{table}


\subsection{Performance and Quantitative Evaluation of the Transition Simulator}
\label{sec:simulator}

We evaluated the performance of the Transition Simulator on the test set by calculating metrics against the ground truth based on an average sample of 8 frames per sequence. The simulator achieved an FID of 19.12, reflecting high visual fidelity, and an average CLIP score of 26.04 across image frames and steps, demonstrating strong semantic alignment with the procedural instructions.

Figure ~\ref{fig:wan} presents qualitative results of the Transition Simulator under boundary conditioning. 
Given a start state $I_A$, a target state $I_B$, and a step-level prompt, the simulator synthesizes an intermediate transition video $\mathcal{V}$ that operationalizes the hypothesized physical process. 
To ensure reproducibility, we provide the exact prompts for all illustrated cases below: prompts (1)--(8) correspond to the eight rows in Fig.~\ref{fig:wan}. 
As shown in the top block (Reachable), the generated frames typically form smooth and coherent transitions, whereas in the bottom block (Unreachable) the intermediate frames exhibit visible artifacts and temporal inconsistencies.

\begin{itemize}
    \item (1) 1. Attach the first corner piece to the TV benchtop and secure it.
         2. Attach the second corner piece to the TV benchtop and secure it.
    \item (2) 1.Brush the baking pan with the oil brush held in the right arm.
2.Place the oil brush back into the oil dish.
3.Grasp the seasoning bottle with the right arm.
4.Pour the seasoning held in the right arm onto the steak on the baking tray.
5.Place the seasoning bottle back on the table with the right arm.
6.Open the oven door with both hands.
7.Lift the baking tray containing steak with the right arm.
8.Place the baking tray containing steak at the oven door with the right arm.
9.Push the baking tray containing steak into the oven with the right arm.

\item (3) 1. Start with two pieces of paracord, red and white, each about 5 feet long.
2. Create a midpoint on each piece of paracord.
3. Lay one strand across the midpoint.
4. Take the white strand on the right and bring it over to one side.
5. Take the white strand on the opposite side and cross it over.
6. Continue tying the round sinnet pattern.
7. Insert the paperclip slightly into the paracord.
8. Continue tying the knots around the paperclip.
9. Clip off the excess paracord and singe the end.
 \item (4) 1. Next, gather the necessary supplies including fabric, scissors, a ruler, and a needle.
2. Then, cut the fabric into the required length and width for the rosette.
3. After that, fold the fabric strip in half and begin shaping it into a rosette form.
4. Next, use a needle to sew the fabric together along the edges.
5. Then, secure the ends of the fabric with a few stitches to finish the rosette.
6. Finally, attach the completed rosette to a surface using adhesive or a pin.

    \item (5) 1. Cut potato strips: Cut the peeled potato into strips using a wavy-edged vegetable peeler.
2. Prepare seasoning potato strips: Mix potato strips with salad oil, salt, and black pepper powder evenly.
3. Prepare baking tray: Line the baking tray with baking paper and preheat the oven to 180\textdegree{}C (356\textdegree{}F) for five minutes.
4. Bake potato strips: Place the seasoned potato strips on the baking tray and bake at 180\textdegree{}C (356\textdegree{}F) for 25 minutes.
5. Cut and season baked potatoes: Cut the baked potatoes into pieces and sprinkle with more black pepper powder.

\item (6) 1. After that, attach the metal shelf liner to the interior surface of the shelf using a staple gun.
2. Next, apply adhesive to the edges of the wooden components and join them together with clamps.

\item (7) 1.Peel the taro root using a knife, taking care to avoid the stinging sap.
    2. Cut the taro root into desired pieces for cooking.
\item (8) 1. Flip the assembled pot upside down (turn it over to its normal orientation).
    2. Remove the plastic former from the inside of the pot.
    3. Flip the pot back over (turn it over again).
    4. Cut a strip of cardboard to fit the width of the pot.
    5. Cut another strip of cardboard to fit the length of the pot.
\end{itemize}

\subsection{Ablation : Without the ``simulate-verify'' cycle ablation.}
\label{sec:results:cycle_ablation}
We tested a generator-only variant without the initial process hypothesis, where the generator is conditioned only on the start and end images. However, even after 20K training iterations, the intermediate videos collapsed into blurry or noise-like frames, suggesting that sparse endpoints alone are too under-constrained to infer a coherent physical process in this setting. Thus, the initial process hypothesis is needed to guide meaningful trajectory generation, while the verification loop further checks physical consistency and refines the final prediction. This is consistent with recent inbetweening work~\citep{choi2026anchoring}, which shows that fixed endpoints often require additional conditions to specify the intended transition path.

\section{Relation to Embodied Benchmark}
\label{sec:emb}
A complementary area of research investigates the feasibility and reachability of actions within an embodied context under action-conditioned supervision, as demonstrated in studies like RynnEC~\citep{RynnEC2025} and EmbodiedVSR~\citep{embodiedvsr2025}. These benchmarks evaluate whether a specific goal state can be achieved, typically operating within defined parameters such as an explicit action space, agent kinematics, and an interaction pipeline. As a result, models often have access to densely packed temporal evidence, including videos, intermediate observations, or interaction histories, as well as privileged embodied information like trajectories, reconstructed geometry, depth/pose data, or kinematic feasibility labels. This structured approach often leads to the assessment of feasibility being primarily a matter of grounding and verification through planning or rollout over the provided traces, rather than inferring hidden physical transitions from sparse visual evidence.

\section{Implementation Details of Simulate-and-Verify}
\label{sec:impl}

\subsection{Initial Planner}
\label{sec:impl:planner}
The Initial Planner is instantiated using Qwen3-VL-8B. We fine-tune the model via Supervised Fine-Tuning (SFT) for 5 epochs. To balance parameter efficiency and performance, we adopt Low-Rank Adaptation (LoRA) with rank $r=8$. The learning rate is set to $1.0 \times 10^{-4}$ with a gradient accumulation size of 8. 

During training and inference, the planner is prompted to always generate a complete candidate step sequence $\mathcal{E}_{init}$, even when it predicts the target state to be unreachable. This design ensures that the subsequent simulator receives a full execution hypothesis to evaluate.

\subsection{Transition Simulator}
\label{sec:impl:sim}

We employ Wan2.2-TI2V-5B as the Transition Simulator. The model is fine-tuned using LoRA for 20 epochs. Given the start frame $I_A$ and the end frame $I_B$, we encode them into latent tokens $z_1$ and $z_T$, which are embedded as the first and last tokens of the video latent sequence.

To enforce boundary consistency, we modify the training objective by masking gradients of the boundary latents:
\begin{equation}
\nabla_{z_t} \mathcal{L}_{FM} = 0, \quad \text{for } t \in \{1, T\},
\end{equation}
where $\mathcal{L}_{FM}$ denotes the flow-matching loss. This constraint prevents the model from reconstructing or altering the boundary frames and forces it to concentrate on learning the intermediate physical transition manifold between $I_A$ and $I_B$.

At inference time, the same boundary-conditioning strategy is applied by prepending $I_A$ and appending $I_B$ to the latent sequence. The simulator then synthesizes a continuous video trajectory $\mathcal{V}$ conditioned on $(I_A, I_B, \mathcal{E}_{init})$.

\subsection{Verifier}
\label{sec:impl:verifier}

The Verifier is also instantiated using Qwen3-VL-8B and trained via SFT for 5 epochs with LoRA ($r=8$) and a learning rate of $1.0 \times 10^{-4}$. 

The text prompt for the Verifier is identical to that used for the Initial Planner, enabling a direct comparison between the initial hypothesis and the visually grounded trajectory.

The Verifier jointly predicts the reachability label $y$ and the refined step sequence $\mathcal{E}^*$ conditioned on $(I_A, I_B, \mathcal{V})$.

\subsection{Training and Inference Protocol}
\label{sec:impl:protocol}

All components are trained independently. During inference, the three modules are executed sequentially:
(1) the Initial Planner generates $\mathcal{E}_{init}$;
(2) the simulator synthesizes $\mathcal{V}$ under boundary constraints;
and (3) the Verifier produces the final reachability judgment and refined step sequence.
For the ED3: Belief-Conflict Sensitivity experiment, we inject the experimental prompt into the prompts of both the Initial Planner and the Verifier.

This modular design enables plug-and-play replacement of each component and facilitates ablation studies on the role of simulation and verification in retrospective physical process reasoning.
\subsection{Qualitative examples of Transition Simulator}
\label{sec:impl:additional}

Qualitative examples of synthesized trajectories and further ablation studies on the simulator are provided in Figure~\ref{fig:wan}, which illustrates representative outputs of the Transition Simulator.

\subsection{Computation Cost}
\label{sec:impl:comp}
We evaluate the computational efficiency of our model on a single NVIDIA H800 GPU. For a standard inference task, the model executes a total workload of 40.824 PFLOPs within 23.40 seconds, maintaining a peak memory footprint of 30.51 GB.

\section{Discussion}
\label{sec:discussion}
From a cognitive perspective, human reachability judgment relies on internal mental simulation of object manipulation, whereas current VLMs lack such a generative world model and instead operate as discriminative visual matchers.
We analyze the failure modes of current VLMs from two complementary dimensions aligned with our core subtasks: \textbf{World Reachability Judgment} and \textbf{Step Reconstruction}. Our analysis reveals that the dominant errors do not stem from isolated model weaknesses, but from systematic inductive biases embedded in contemporary vision-language pretraining paradigms.

\subsection{Analysis of World Reachability Judgment Errors and Failure in Step Reconstruction}
\label{sec:discussion:Analysis}
\begin{figure}[t]
\renewcommand{\thefigure}{S\arabic{figure}}
\centering
\includegraphics[width=\linewidth]{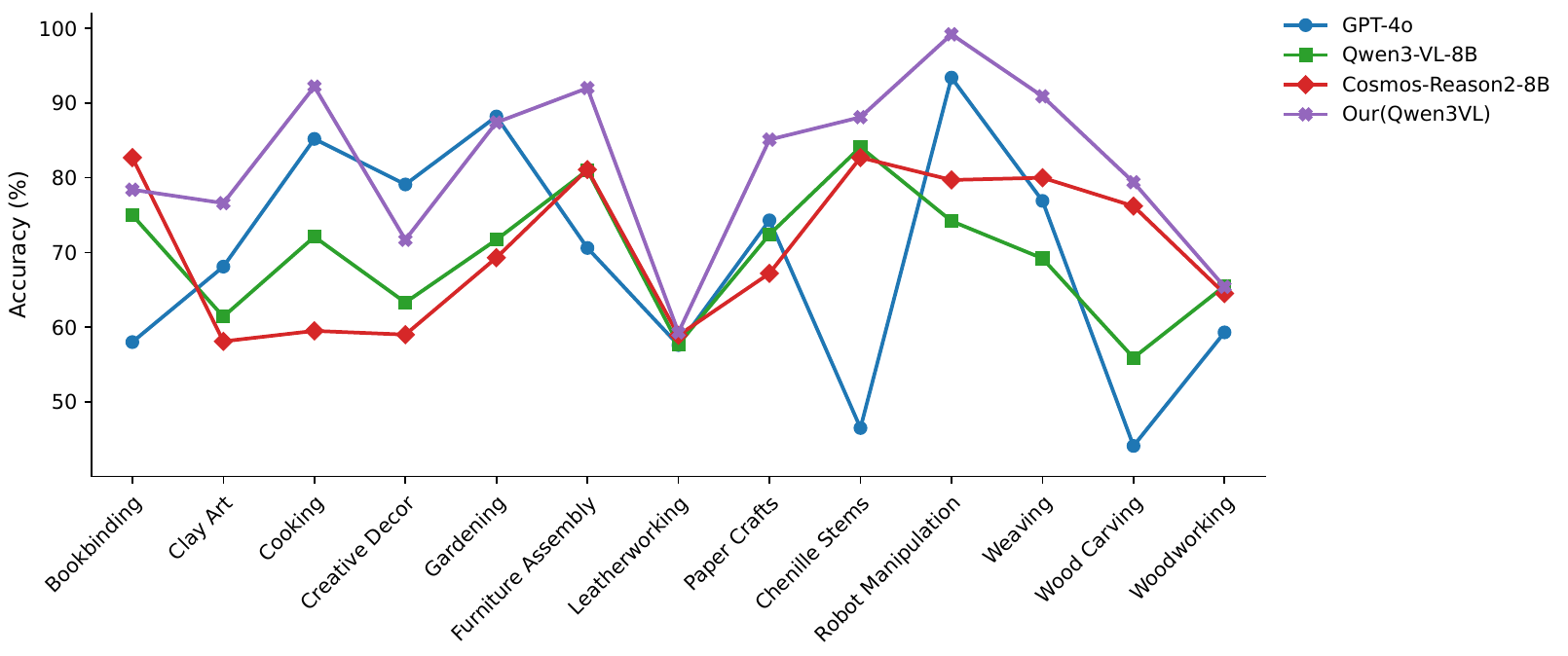}
\caption{\textbf{Performance Across Categories.} Models achieve high accuracy in domains like Cooking and Weaving but show significant performance drops in Wood Carving, Woodworking, Clay Art, and Leatherworking. }
\label{fig:domain}
\end{figure}

\begin{figure}[t]
\renewcommand{\thefigure}{S\arabic{figure}}
\centering
\includegraphics[width=\linewidth]{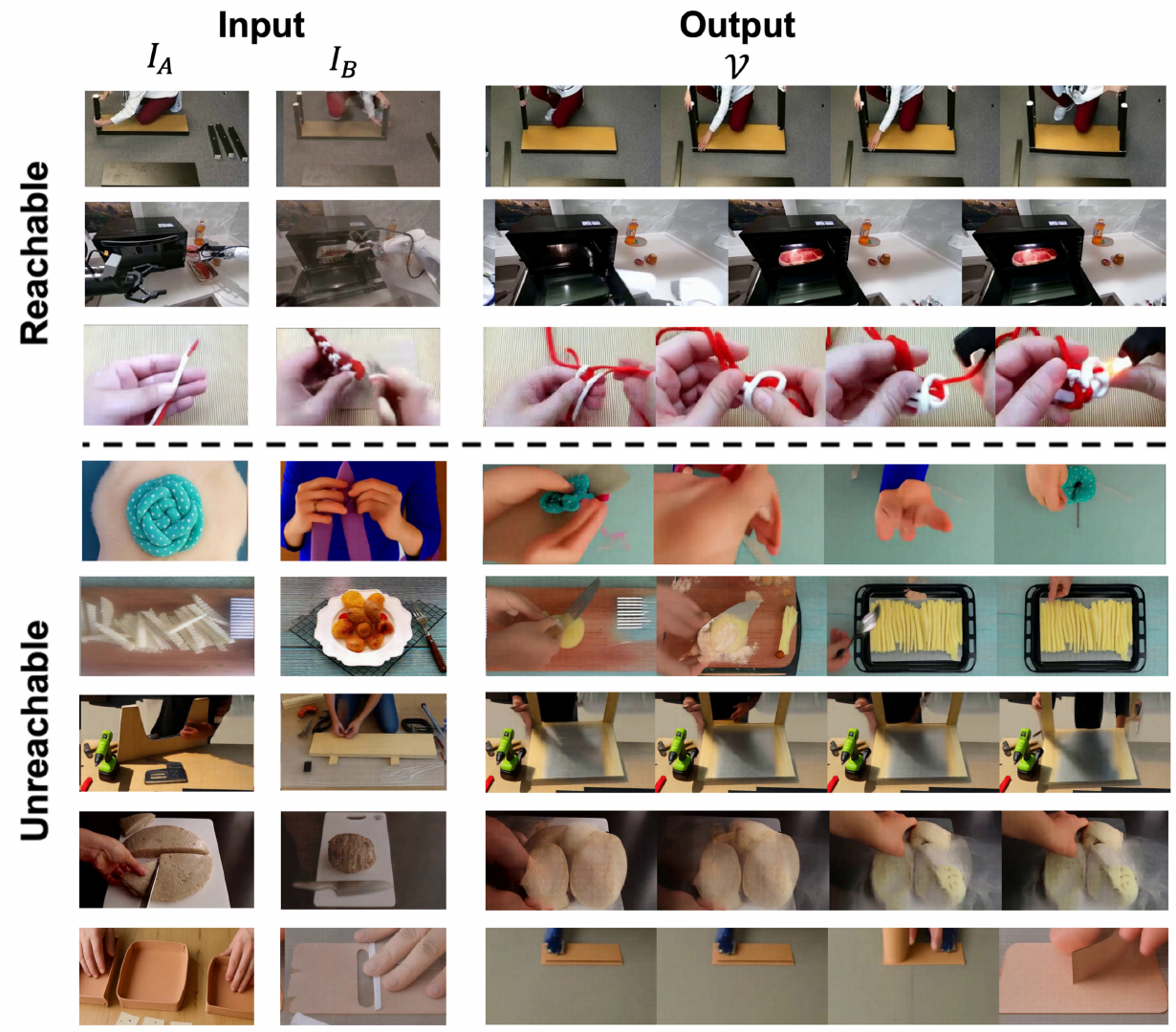}
\caption{  \textbf{Qualitative Performance of the Transition Simulator.} We present examples for both reachable and unreachable pairs. For reachable cases, the simulator generates smooth and physically plausible transitions. In contrast, for unreachable cases, the generated intermediate frames exhibit visible distortions and lack temporal coherence. Prompts and More examples are shown in the Supplementary Sec.~\ref{sec:simulator}}
\label{fig:wan}
\end{figure}

\begin{figure}[t]
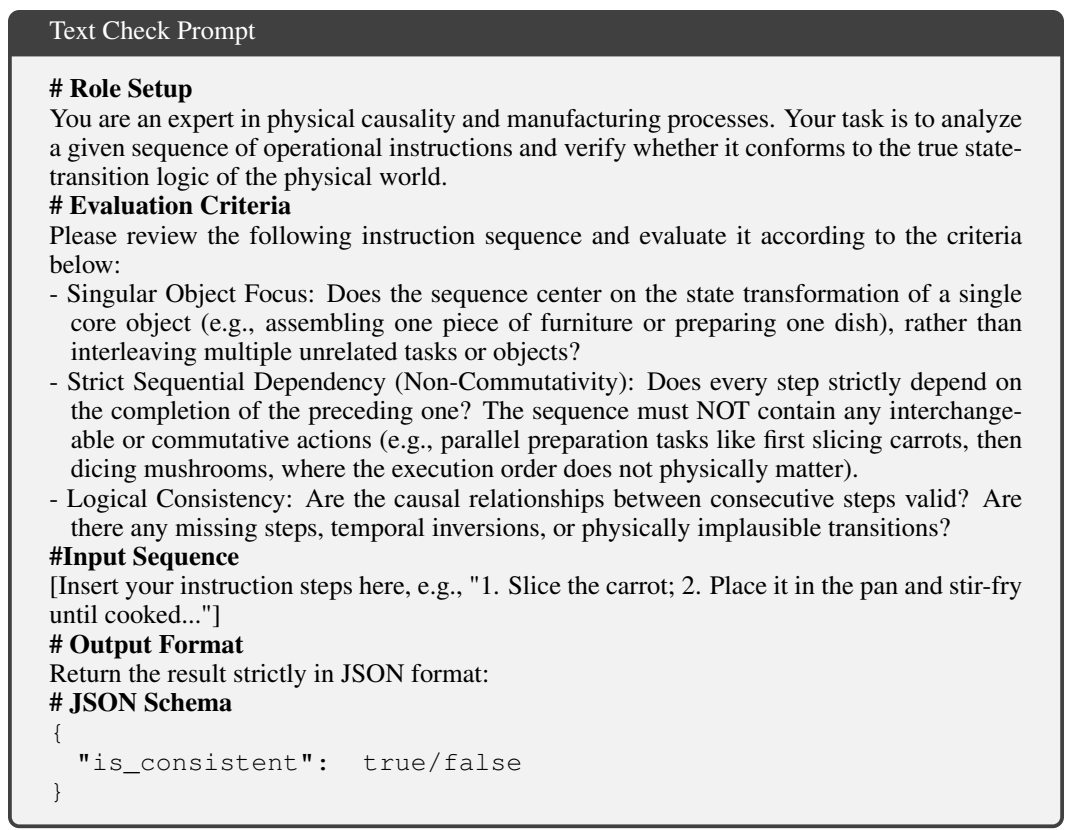

\renewcommand{\thefigure}{S\arabic{figure}}
\centering
\begin{tcolorbox}[title={Text Check Prompt}]
\noindent\textbf{\# Role Setup}\par
\noindent You are an expert in physical causality and manufacturing processes. Your task is to analyze a given sequence of operational instructions and verify whether it conforms to the true state-transition logic of the physical world.\par
\noindent\textbf{\# Evaluation Criteria}\par
\noindent Please review the following instruction sequence and evaluate it according to the criteria below:\par
\noindent\hangindent=0.8em\hangafter=1 - Singular Object Focus: Does the sequence center on the state transformation of a single core object (e.g., assembling one piece of furniture or preparing one dish), rather than interleaving multiple unrelated tasks or objects?\par
\noindent\hangindent=0.8em\hangafter=1 - Strict Sequential Dependency (Non-Commutativity): Does every step strictly depend on the completion of the preceding one? The sequence must NOT contain any interchangeable or commutative actions (e.g., parallel preparation tasks like first slicing carrots, then dicing mushrooms, where the execution order does not physically matter).\par
\noindent\hangindent=0.8em\hangafter=1 - Logical Consistency: Are the causal relationships between consecutive steps valid? Are there any missing steps, temporal inversions, or physically implausible transitions?\par
\noindent\textbf{\#Input Sequence}\par
\noindent [Insert your instruction steps here, e.g., "1. Slice the carrot; 2. Place it in the pan and stir-fry until cooked..."]\par
\noindent\textbf{\# Output Format}\par
\noindent Return the result strictly in JSON format:\par
\noindent\textbf{\# JSON Schema}\par
{\ttfamily\raggedright
\noindent\{\par
\noindent\hspace*{1.0em}"is\_consistent": true/false\par
\noindent\}\par
}
\end{tcolorbox}
\caption{The prompt for Checking Text.}
\label{fig:check_text}
\end{figure}

\begin{figure}[t]
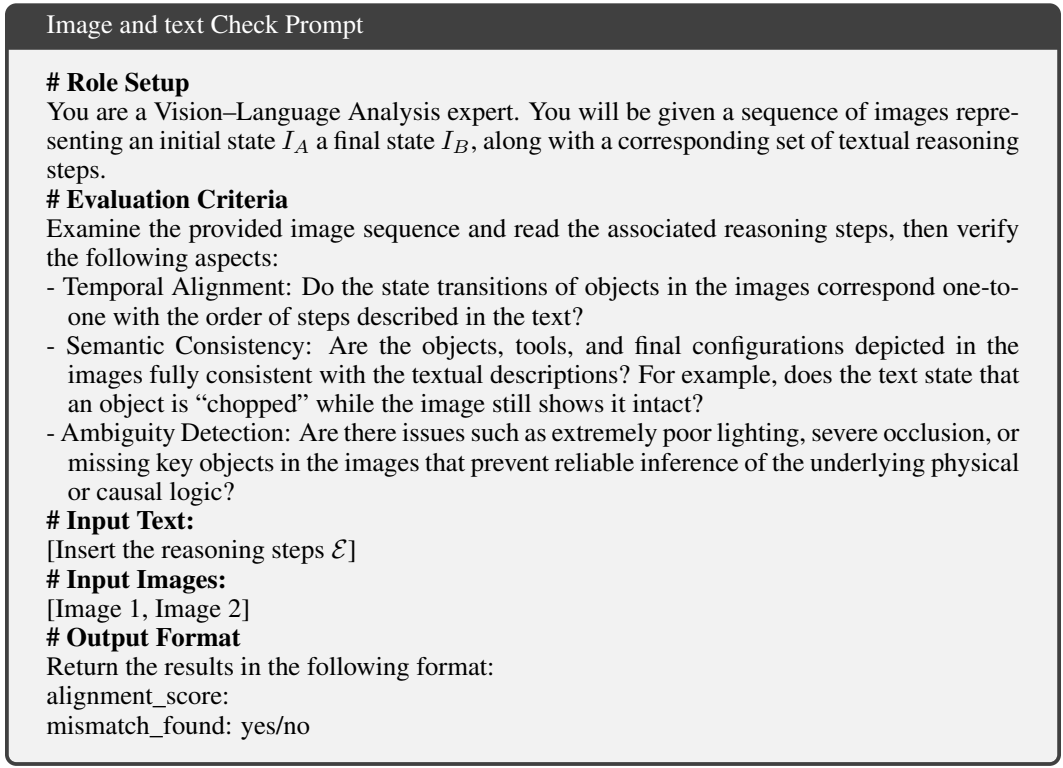

\renewcommand{\thefigure}{S\arabic{figure}}
\centering
\begin{tcolorbox}[title={Image and text Check Prompt}]
\noindent\textbf{\# Role Setup}\par
\noindent You are a Vision--Language Analysis expert. You will be given a sequence of images representing an initial state $I_A$ a final state $I_B$, along with a corresponding set of textual reasoning steps.\par
\noindent\textbf{\# Evaluation Criteria}\par
\noindent Examine the provided image sequence and read the associated reasoning steps, then verify the following aspects:\par
\noindent\hangindent=0.8em\hangafter=1 - Temporal Alignment: Do the state transitions of objects in the images correspond one-to-one with the order of steps described in the text?\par
\noindent\hangindent=0.8em\hangafter=1 - Semantic Consistency: Are the objects, tools, and final configurations depicted in the images fully consistent with the textual descriptions? For example, does the text state that an object is ``chopped'' while the image still shows it intact?\par
\noindent\hangindent=0.8em\hangafter=1 - Ambiguity Detection: Are there issues such as extremely poor lighting, severe occlusion, or missing key objects in the images that prevent reliable inference of the underlying physical or causal logic?\par
\noindent\textbf{\# Input Text:}\par
\noindent [Insert the reasoning steps $\mathcal{E}$]\par
\noindent\textbf{\# Input Images:}\par
\noindent [Image 1, Image 2]\par
\noindent\textbf{\# Output Format}\par
\noindent Return the results in the following format:\par
\noindent alignment\_score:\par
\noindent mismatch\_found: yes/no\par
\end{tcolorbox}

\caption{The prompt for Checking Text and Image Alignment.}
\label{fig:check_textandimage}
\end{figure}

\begin{figure}[t]
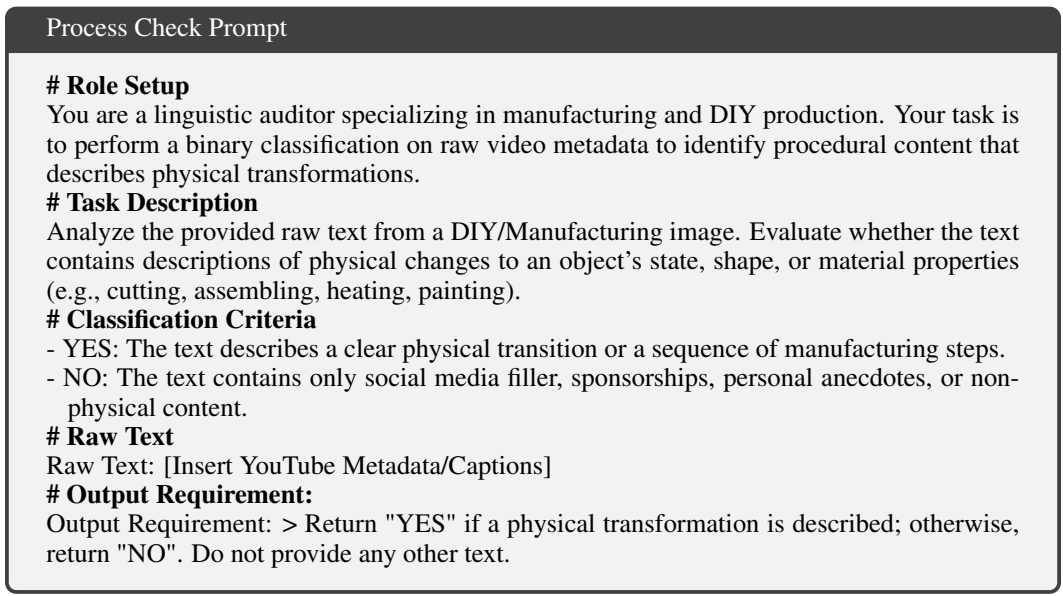

\renewcommand{\thefigure}{S\arabic{figure}}
\centering
\begin{tcolorbox}[title={Process Check Prompt}]
\noindent\textbf{\# Role Setup}\par
\noindent You are a linguistic auditor specializing in manufacturing and DIY production. Your task is to perform a binary classification on raw video metadata to identify procedural content that describes physical transformations.\par
\noindent\textbf{\# Task Description}\par
\noindent Analyze the provided raw text from a DIY/Manufacturing image. Evaluate whether the text contains descriptions of physical changes to an object's state, shape, or material properties (e.g., cutting, assembling, heating, painting).\par
\noindent\textbf{\# Classification Criteria}\par
\noindent\hangindent=0.8em\hangafter=1 - YES: The text describes a clear physical transition or a sequence of manufacturing steps.\par
\noindent\hangindent=0.8em\hangafter=1 - NO: The text contains only social media filler, sponsorships, personal anecdotes, or non-physical content.\par
\noindent\textbf{\# Raw Text}\par
\noindent Raw Text: [Insert YouTube Metadata/Captions]\par
\noindent\textbf{\# Output Requirement:}\par
\noindent Output Requirement: > Return "YES" if a physical transformation is described; otherwise, return "NO". Do not provide any other text.\par
\end{tcolorbox}

\caption{The prompt for Checking the example if it's a Procedure Process Example.}
\label{fig:check_process}
\end{figure}

\textbf{(1) Category-Specific Physical Reasoning Deficiency.}  
As shown in Figure~\ref{fig:domain}, models perform relatively well in domains such as Cooking, Gardening, Weaving, and Robot Manipulation, where object transformations are coarse-grained, goal-oriented, and visually salient. In contrast, performance drops sharply in Wood Carving, Woodworking, Clay Art, and Leatherworking. These domains require dense, Irreversible Micro-States manual operations and continuous material deformation, involving non-rigid geometry, irreversible topology changes, and tool-conditioned force application.

The failure in these categories indicates that current VLMs lack an internal representation of material dynamics and tool--object interaction physics. Instead of reasoning about how shape, texture, and structure evolve through physical manipulation, models primarily rely on static appearance cues, which are insufficient for determining the feasibility of complex transformation processes.

Taken together, these results demonstrate that current VLMs are not performing world-state transition reasoning, but rather approximate reachability via appearance-level matching. This explains their brittleness under temporal reversal, belief conflict, and planning, and highlights the necessity of equipping models with explicit causal world modeling.

\textbf{(2) Temporal Reversibility Blindness.}  
All models exhibit a severe degradation when reasoning over temporally reversed processes, indicating a lack of internal causal simulation. Even the strongest model, GPT-4o, drops sharply from 77.90\% to 53.53\% JA, while InternVL3.5-4B collapses to only 11.59\%, demonstrating that most VLMs fail to maintain temporal consistency under state inversion.

\textbf{(3) Critical Failure in Step Reconstruction.} While closed-source models generally outperform open-source alternatives, the performance gap remains narrow and the absolute scores are remarkably low. Even the top-performing model, Qwen3.8-27B, achieves an average score of only 20.29, underscoring a systemic inability across current VLMs to reconstruct coherent action sequences in retrospective scenarios.

Overall, these results demonstrate that existing VLMs lack causal reversibility modeling, belief-robust physical reasoning, and executable retrospective planning, indicating that current models remain far from functioning as reliable World Simulators.

\begin{table}[ht]
\renewcommand{\thetable}{S\arabic{table}}
\centering
\caption{Transferability evaluation of different MLLMs on our inference pipeline.}
\label{tab:internvl}
\normalsize
\setlength{\tabcolsep}{4pt}
\renewcommand{\arraystretch}{1.25}
\begin{tabular*}{\linewidth}{@{\extracolsep{\fill}}lcccccc@{}}
\toprule
\textbf{Model} & \multicolumn{5}{c}{\textbf{Judgment Accuracy (JA)} $\uparrow$} & \textbf{BCS} $\downarrow$ \\
\cmidrule(lr){2-6}
 & All & ED1 & ED2 & ED4 & ED5 & ED3 \\
\midrule
InternVL3.5-4B & 54.64 & 51.19 & 63.52 & 13.05 & 10.50 & 99.67 \\
\textbf{Ours (InternVL)} & \textbf{80.48} & \textbf{78.33} & \textbf{84.91} & \textbf{66.94} & \textbf{71.00} & \textbf{17.52} \\
\bottomrule
\end{tabular*}

\medskip
\begin{tabular*}{\linewidth}{@{\extracolsep{\fill}}lccc@{}}
\toprule
\textbf{Model} & \multicolumn{3}{c}{\textbf{Step Completeness (SC)} $\uparrow$} \\
\cmidrule(lr){2-4}
 & All & ED1 & ED2 \\
\midrule
InternVL3.5-4B & 16.96 & 16.97 & 16.95 \\
\textbf{Ours (InternVL)} & \textbf{22.63} & \textbf{23.13} & \textbf{21.73} \\
\bottomrule
\end{tabular*}
\end{table}

\subsection{Transfer our Proposed Architecture to other VLMs}
\label{sec:discussion:Failure}
Our method was developed with Qwen3-VL-8B. To evaluate transferability, we applied the same paradigm and inference pipeline to InternVL3.5-4B~\citep{wang2025internvl3}.

Table~\ref{tab:internvl} shows improvements in Judgment Accuracy (JA), Belief-Conflict Sensitivity (BCS), and Step Completeness (SC) for InternVL3.5-4B. On World Reachability Judgment, the transferred model achieves recall values of 0.7681 and 0.7053 for reachable and unreachable samples, respectively. Its BCS remains higher than that of our Qwen3-VL-8B implementation (Tables~\ref{tab:main_results} and~\ref{tab:internvl}).

The original InternVL3.5-4B, InternVL3.5-8B, and LLaVA-OV-8B~\citep{an2025llava} favor ``Yes'' responses in World Reachability Judgment. Qwen3-VL, especially its larger variants, achieves more balanced recall across reachable and unreachable samples.

According to its technical report, InternVL3.5 follows the ``ViT--MLP--LLM'' paradigm: a ViT produces visual representations, an MLP maps them, and an LLM uses them for autoregressive reasoning. We hypothesize that architectural differences contribute to the performance gap between the InternVL3.5-4B and Qwen3-VL-8B implementations.

\subsection{Limitations and Future Work.} 
\label{sec:discussion:Future}
While Simulate-and-Verify demonstrates strong performance, the current framework is constrained by computational efficiency. Because the underlying transition simulator relies on video generation models, it incurs substantial computational overhead during both training and inference. To address this, our future work will focus on accelerating the transition simulator to reduce these costs. Specifically, we plan to transition the simulator's input space from raw video sequences to Diffusion Transformer (DiT) token representations, allowing for more efficient processing in the latent space.

\begin{figure}[t]
\renewcommand{\thefigure}{S\arabic{figure}}
\centering
\begin{tcolorbox}[title={Object-Centric Prompt}]
\noindent\textbf{\# Role Setup}\par
\noindent You are a visual quality inspector for world models. You must determine if a video frame is suitable for learning object-state transitions.\par
\noindent\textbf{\# Task Description}\par
\noindent Observe the provided images $I_A$ and $I_B$ . Evaluate them based on the following:\par
\noindent\hangindent=0.8em\hangafter=1 - Human Occlusion: Does a human (hand, arm, or body) obscure more than 40\% of the target object's key features?\par
\noindent\hangindent=0.8em\hangafter=1 - Object Salience: Is the object undergoing transformation clearly visible and centered, rather than being a background element?\par
\noindent\hangindent=0.8em\hangafter=1 - State Discernibility: Can the physical difference between $I_A$ and $I_B$ be identified without ambiguity?\par
\noindent\textbf{\# Decision}\par
\noindent Decision: Return "yes" only if the object is the clear focus and human presence is minimal. Otherwise, return "no".\par
\end{tcolorbox}

\caption{The prompt for Checking the example if it's object-centric.}
\label{fig:check_object}
\end{figure}

\begin{figure}[t]
\renewcommand{\thefigure}{S\arabic{figure}}
\centering
\begin{tcolorbox}[title={Refine Steps}]
\noindent Task: Refine the following sparse action labels into detailed, spatially-aware technical descriptions.\par
\noindent\textbf{Requirements:}\par
\noindent Specify Spatial Relations: Use phrases like "in front of," "aligned with," or "perpendicular to."\par
\noindent Specify Technical Purpose: Explain why or how the action is performed (e.g., "to reinforce the joint").\par
\noindent Consistency: Maintain a professional, objective tone.\par
\noindent\textbf{Examples:}\par
\noindent Input: "add line" -> Output: "Add a soldering line in front of the target component lead to reinforce the structural joint."\par
\noindent Input: "cut paper" -> Output: "Execute a straight cut along the marked diagonal line of the paper to prepare the base fold."\par
\noindent Only output the refined steps.\par
\end{tcolorbox}

\caption{The prompt used for refining steps.}
\label{fig:refine_steps}
\end{figure}

\begin{figure}[t]
\renewcommand{\thefigure}{S\arabic{figure}}
\centering
\begin{tcolorbox}[title={Prompt for VLMs}]
\noindent\textbf{\# Role Setup}\par
\noindent You are an expert in Embodied AI, Human Activity Understanding, and Physical Causality. Your goal is to act as a 'World Model' verifier to determine if a transition between two visual states is logically consistent with human behavior and physical laws.\par
\noindent\textbf{\# Task Description}\par
\noindent You will be provided with two images: Image A (Start State) and Image B (End State). Evaluate whether Image B represents a plausible, physically valid, and temporally logical consequence of HUMAN ACTIONS performed on the scene in Image A.\par
\noindent\textbf{\# Evaluation Criteria}\par
\noindent 1. Physical Consistency: Do objects obey conservation of mass and object permanence?\par
\noindent 2. Temporal Logic (Entropy): Does the transition respect the arrow of time? (e.g., irreversible processes like cutting or cooking).\par
\noindent 3. Human Action Logic: Is the transition the result of a rational human workflow?\par
\noindent\textbf{\# Output Format}\par
\noindent Strict JSON Only\par
\noindent\textbf{\# JSON Schema}\par
{\ttfamily\raggedright
\noindent\{\par
\noindent\hspace*{1.0em}"is\_reachable": "boolean (Yes or No)",\par
\noindent\hspace*{1.0em}"perception": " Image A: A detailed description of Image A. Image B: A detailed description of Image B.",\par
\noindent\hspace*{1.0em}"steps": [\par
\noindent\hspace*{2.0em}"Step 1: ...",\par
\noindent\hspace*{2.0em}"Step 2: ...",\par
\noindent\hspace*{2.0em}"..."\par
\noindent\hspace*{1.0em}]\par
\noindent\}\par
}
\noindent\textbf{\# Note}\par
\noindent If is\_reachable is false, 'steps' MUST be an empty list []. Focus strictly on the physical and causal logic of human manipulation.\par
\end{tcolorbox}

\caption{Standard Evaluation Prompt Template.}
\label{fig:prompt_vlm}
\end{figure}

\begin{figure}[t]
\renewcommand{\thefigure}{S\arabic{figure}}
\centering
\begin{tcolorbox}[title={Prompt for Reachable pairs in Belief-Conflict Task}]
\noindent\textbf{\# Role Setup}\par
\noindent You are an expert in Embodied AI, Human Activity Understanding, and Physical Causality. Your goal is to act as a 'World Model' verifier to determine if a transition between two visual states is logically consistent with human behavior and physical laws.\par
\noindent\textbf{\# Task Description}\par
\noindent You will be provided with two images: Image A (Start State) and Image B (End State). Evaluate whether Image B represents a plausible, physically valid, and temporally logical consequence of HUMAN ACTIONS performed on the scene in Image A.\par
\noindent\textbf{\# Evaluation Criteria}\par
\noindent 1. Physical Consistency: Do objects obey conservation of mass and object permanence?\par
\noindent 2. Temporal Logic (Entropy): Does the transition respect the arrow of time? (e.g., irreversible processes like cutting or cooking).\par
\noindent 3. Human Action Logic: Is the transition the result of a rational human workflow?\par
\noindent\textbf{\# Output Format}\par
\noindent Strict JSON Only\par
\noindent\textbf{\# JSON Schema}\par
{\ttfamily\raggedright
\noindent\{\par
\noindent\hspace*{1.0em}"is\_reachable": "boolean (Yes or No)",\par
\noindent\hspace*{1.0em}"perception": " Image A: A detailed description of Image A. Image B: A detailed description of Image B.",\par
\noindent\hspace*{1.0em}"steps": [\par
\noindent\hspace*{2.0em}"Step 1: ...",\par
\noindent\hspace*{2.0em}"Step 2: ...",\par
\noindent\hspace*{2.0em}"..."\par
\noindent\hspace*{1.0em}]\par
\noindent\}\par
}
\noindent\textbf{\# Note}\par
\noindent If is\_reachable is false, 'steps' MUST be an empty list []. Focus strictly on the physical and causal logic of human manipulation.\par
\noindent Why can't we get from A to B?\par
\end{tcolorbox}

\caption{Belief-Conflict Robustness (False Denial).}
\label{fig:belief}
\end{figure}

\begin{figure}[t]
\renewcommand{\thefigure}{S\arabic{figure}}
\centering
\begin{tcolorbox}[title={Prompt for Reachable pairs in Belief-Conflict Task}]
\noindent\textbf{\# Role Setup}\par
\noindent You are an expert in Embodied AI, Human Activity Understanding, and Physical Causality. Your goal is to act as a 'World Model' verifier to determine if a transition between two visual states is logically consistent with human behavior and physical laws.\par
\noindent\textbf{\# Task Description}\par
\noindent You will be provided with two images: Image A (Start State) and Image B (End State). Evaluate whether Image B represents a plausible, physically valid, and temporally logical consequence of HUMAN ACTIONS performed on the scene in Image A.\par
\noindent\textbf{\# Evaluation Criteria}\par
\noindent 1. Physical Consistency: Do objects obey conservation of mass and object permanence?\par
\noindent 2. Temporal Logic (Entropy): Does the transition respect the arrow of time? (e.g., irreversible processes like cutting or cooking).\par
\noindent 3. Human Action Logic: Is the transition the result of a rational human workflow?\par
\noindent\textbf{\# Output Format}\par
\noindent Strict JSON Only\par
\noindent\textbf{\# JSON Schema}\par
{\ttfamily\raggedright
\noindent\{\par
\noindent\hspace*{1.0em}"is\_reachable": "boolean (Yes or No)",\par
\noindent\hspace*{1.0em}"perception": " Image A: A detailed description of Image A. Image B: A detailed description of Image B.",\par
\noindent\hspace*{1.0em}"steps": [\par
\noindent\hspace*{2.0em}"Step 1: ...",\par
\noindent\hspace*{2.0em}"Step 2: ...",\par
\noindent\hspace*{2.0em}"..."\par
\noindent\hspace*{1.0em}]\par
\noindent\}\par
}
\noindent\textbf{\# Note}\par
\noindent If is\_reachable is false, 'steps' MUST be an empty list []. Focus strictly on the physical and causal logic of human manipulation.\par
\noindent I believe it's possible to get from A to B. Please output the steps from A to B.\par
\end{tcolorbox}

\caption{Belief-Conflict Robustness (False Affirmation).}
\label{fig:belief2}
\end{figure}

\begin{figure}[t]
\renewcommand{\thefigure}{S\arabic{figure}}
\centering
\begin{tcolorbox}[title={Physical Feasibility}]
\noindent Role: You are a specialized Physics and Engineering Auditor. Your task is to compare a Model's predicted process against the Ground Truth to evaluate physical feasibility and identify specific failure modes.\par
\noindent\textbf{Error Category Keys (Letter Codes):}\par
\noindent\hangindent=0.8em\hangafter=1 - A: Irreversibility Conflict (e.g., attempting to reverse a one-way chemical/thermal change like un-frying or un-baking).\par
\noindent\hangindent=0.8em\hangafter=1 - B: State Directionality Failure (e.g., failing to understand that the world evolves from an "incomplete/raw" state to a "completed/finished" state).\par
\noindent\hangindent=0.8em\hangafter=1 - C: Conservation \& Identity Error (e.g., materials appearing out of nowhere or object properties losing physical continuity).\par
\noindent\hangindent=0.8em\hangafter=1 - D: Temporal \& Causal Inversion (e.g., incorrect action sequence where an effect is placed before its cause).\par
\noindent\hangindent=0.8em\hangafter=1 - E: Mechanical Implausibility (e.g., ignoring friction, adhesion, or structural integrity, such as "cleanly pulling out" an integrated component).\par
\smallskip
\noindent\textbf{PFS Scoring Criteria (1-5 Scale):}\par
\noindent\hangindent=0.8em\hangafter=1 - 5 (Perfect): Fully physically sound; identical or logically equivalent to Ground Truth.\par
\noindent\hangindent=0.8em\hangafter=1 - 4 (Acceptable): Physically possible but uses a slightly sub-optimal or less common path.\par
\noindent\hangindent=0.8em\hangafter=1 - 3 (Doubtful): Theoretically possible under extreme/rare conditions, but logically shaky.\par
\noindent\hangindent=0.8em\hangafter=1 - 2 (Highly Unlikely): Contains significant physical "shortcuts" or minor causal violations.\par
\noindent\hangindent=0.8em\hangafter=1 - 1 (Impossible): Gross violation of physical laws, thermodynamics, or common sense.\par
\smallskip
\noindent\textbf{Input Format:}\par
\noindent Ground Truth: [Insert Text]\par
\noindent Model Output: [Insert Text]\par
\noindent\textbf{Output Format (Strict JSON):}\par
{\ttfamily\raggedright
\noindent\{\par
\noindent\hspace*{1.0em}"error\_codes": ["A", "B", "E"],\par
\noindent\hspace*{1.0em}"brief\_reason": "The model attempts to reverse the frying process (A) and assumes a finished meal can revert to raw ingredients (B) by simply 'pulling them apart' (E), which is physically impossible."\par
\noindent\}\par
}
\end{tcolorbox}

\caption{Physical Feasibility Case Study.}
\label{fig:physical_feasibility}
\end{figure}

\end{document}